\documentclass[pdflatex,sn-basic]{sn-jnl}

\usepackage{graphicx}%
\usepackage{subcaption}%
\usepackage{multirow}%
\usepackage{amsmath,amssymb,amsfonts}%
\usepackage{mathrsfs}%
\usepackage[title]{appendix}%
\usepackage[table]{xcolor}%
\usepackage{textcomp}%
\usepackage{manyfoot}%
\usepackage{booktabs}%
\usepackage{algorithm}%
\usepackage{algorithmicx}%
\usepackage{algpseudocode}%
\usepackage{listings}%
\usepackage{tcolorbox}%
\def\BibTeX{{\rm B\kern-.05em{\sc i\kern-.025em b}\kern-.08em
    T\kern-.1667em\lower.7ex\hbox{E}\kern-.125emX}}

\usepackage{cite}
\graphicspath{{fig/}}
\DeclareGraphicsExtensions{.pdf}
\usepackage{url}

\usepackage{hyperref}
\makeatletter
\def\toclevel@paragraph{3}
\makeatother
\usepackage[T1]{fontenc}

\usepackage{tikz}
\usetikzlibrary{chains}
\usetikzlibrary{arrows.meta}
\usetikzlibrary{patterns}

\usepackage[bottom]{footmisc}

\newcommand{\eg}{\textit{e}.\textit{g}., }
\newcommand{\ie}{\textit{i}.\textit{e}., }
\newcommand{\etal}{\textit{et al}.}
\newcommand{\etc}{\textit{etc}.}
\newcommand{\code}[1]{\texttt{#1}}
\newcounter{number}
\newcommand{\finding}[1]{%
  \stepcounter{number}%
  \begin{tcolorbox}[
    colback=gray!10,
    colframe=gray!40,
    left=3pt,
    right=3pt,
    top=3pt,
    bottom=3pt,
  ]
  \textbf{Finding {\thenumber}} \emph{#1}
  \end{tcolorbox}
}

\makeatletter
\renewcommand\paragraph{\@startsection{paragraph}{4}%
  {\parindent}%
  {1ex}%
  {-0.5em}%
  {\normalfont\normalsize\bfseries}}
\makeatother

\begin{document}

\title{
The Fools are Certain; the Wise are Doubtful:
Exploring LLM Confidence in Code Completion
}



\author*[1]{\fnm{Zoe} \sur{Kotti}}\email{zoekotti@aueb.gr}

\author[1]{\fnm{Konstantina} \sur{Dritsa}}\email{dritsakon@aueb.gr}

\author[1]{\fnm{Diomidis} \sur{Spinellis}}\email{dds@aueb.gr}

\author[1]{\fnm{Panos} \sur{Louridas}}\email{louridas@aueb.gr}

\affil*[1]{%
\orgdiv{Department of Management Science and Technology},
\orgname{Athens University of Economics and Business},
\orgaddress{\city{Athens}, \country{Greece}}%
}


\abstract{ Code completion entails the task of providing missing
  tokens given a surrounding context. It can boost developer
  productivity and serve as a code discovery tool. Code completion has
  recently been approached with Large Language Models (LLMs) fine-tuned
  on code (code LLMs). The performance of code LLMs can be assessed with
  downstream and intrinsic metrics. Downstream metrics are usually employed to
  evaluate the practical utility of a model, but can be unreliable and
  require complex calculations and domain-specific knowledge. In
  contrast, intrinsic metrics such as perplexity, entropy, and mutual
  information, which measure model confidence or uncertainty, are
  simple, versatile, and universal across LLMs and tasks, and have been
  proposed as proxies for functional correctness and hallucination risk
  in LLM-generated code. Motivated by this, we evaluate the confidence
  of LLMs when generating code by measuring code perplexity across
  programming languages, models, and datasets using various LLMs, and
  a sample of 2254 files from 881 GitHub projects. We find that
  strongly-typed languages exhibit lower perplexity than dynamically
  typed languages. Scripting languages also demonstrate higher
  perplexity. Shell appears universally high in perplexity, whereas
  Java appears low. Code perplexity depends on the employed LLM; under
  a fixed model, relative language-level rankings are largely
  stable across evaluation corpora. Although code comments modestly
  increase perplexity, the language ranking based on perplexity is
  barely affected by their presence. LLM researchers, developers, and
  users can use our findings to assess the suitability of LLM-based
  code completion in specific software projects based on how language,
  model choice, and code characteristics impact model confidence. }



\maketitle

\begin{quote}
\textit{
``The fundamental cause of the trouble is that in the modern world
the stupid are cocksure while the intelligent are full of doubt.''
}
\end{quote}
\hfill --- Bertrand Russell~\citeyearpar{russell1998essays}

\section{Introduction}
\label{sec:intro}

Code completion entails the task of providing missing tokens
given a surrounding context~\citep{kim2021code}.
It is commonly encountered in IDEs and powerful editors,
where it provides developers with code suggestions
based on the code they have already typed~\citep{wang2023review}.
Code completion can boost developer productivity
by generating entire code blocks or functions
(with varying degrees of correctness)~\citep{kim2021code, barke2022grounded, li2024acecoder, husein2025llms}.
At the same time,
it makes an unobtrusive code discovery tool
by allowing developers to stay in the same mental context
when looking for some functionality,
instead of browsing through developer question-and-answer websites~\citep{kim2021code, wang2023review}.

Similar to word prediction, Large Language Models (LLMs) can be leveraged
to predict code tokens based on semantic insights from existing code and comments,
provided they can be fine-tuned for code tasks using code data.
This has led to the development of specialized models for code
known as \emph{code LLMs}~\citep{pan2024lost} or \emph{LLMs4Code}~\citep{sharma2024llms}.
Code LLMs find applications in diverse software engineering tasks including
code generation~\citep{zan2023survey},
code understanding~\citep{nam2024using}, and
code translation~\citep{pan2024lost}.

The performance of code LLMs can be assessed with two types of metrics:
downstream and intrinsic.
The former are usually employed to evaluate the practical utility of a model,
while the latter can assess model confidence or uncertainty.
Downstream metrics may not always be reliable,
and typically require domain-specific knowledge and complex calculations~\citep{huyen2019evaluation, liu2024evalplus}.
In contrast,
intrinsic metrics are simple, versatile, and universal
across LLMs and tasks~\citep{huyen2019evaluation, farquhar2024detecting}.
At the same time,
they provide valuable insights during the model training process
by helping the understanding of how effectively the model is learning
to make predictions from the training data~\citep{huyen2019evaluation, li2023starcoder}.
Such insights can be used to adjust the training strategy and model architecture.

Despite these advantages,
downstream metrics dominate current practice in code generation evaluation.
A recent review~\citep{hou2024large} found over 300 papers using 18 different downstream metrics including
\emph{BLEU~Score},
\emph{Pass@k},
\emph{Accuracy/Accuracy@k},
\emph{Exact~Match}, and
\emph{CodeBLEU}.
Only one study~\citep{xu2022polycoder} employed perplexity~\citep{jelinek:1977} as a primary intrinsic metric,
highlighting an opportunity to assess LLM confidence more systematically.

This evaluation gap is further underscored by a growing focus on code hallucinations:
plausible-looking but incorrect code fragments that can mislead developers.
Researchers have recently introduced a taxonomy of hallucinations in code generation,
and a benchmark to detect such issues~\citep{agarwal2024codemirage}.
Complementary to hallucination detection,
understanding and quantifying model uncertainty (and by extension, \emph{confidence})
may offer early indicators of potential issues in generated code
before they are propagated into codebases~\citep{spiess2025calibration}.

Confidence evaluation metrics (\eg entropy, mutual information, perplexity)
can serve as \emph{proxies} for correctness and hallucination risk in LLM-generated code.
Sharma \etal~\citeyearpar{sharma2025assessing} empirically showed that
higher entropy and mutual information correlate with lower functional
correctness.

In this work we focus on \emph{perplexity}. Perplexity measures a
model's confidence in predicting the next token based on the context
provided by the previous tokens~\citep{huyen2019evaluation,
  yang2024unveiling}. The higher the perplexity, the less confident a
model is: intuitively, the more perplexed a model is, the more
uncertain it is about its predictions. Perplexity has been found to be
positively correlated with error rate~\citep{bahl1983likelihood}. In
certain tasks (\eg sentiment analysis, multi-genre natural language
inference, code completion, program synthesis) improved (\ie
lower) perplexity seems to improve end-task performance, thus
providing for practitioners a lightweight and early indicator of
potential issues in generated code~\citep{liu2019roberta,
  xu2022polycoder, nijkamp2023codegen}. As a consequence, perplexity,
as a confidence measure, can offer valuable support for risk
assessment and quality enhancement within development workflows. For
example, development teams could adopt a code review policy based on
the confidence of the LLM-generated code: highly confident code
suggestions could be reviewed more lightly than moderate- or
low-confidence ones~\citep{spiess2025calibration}.

Code produced by LLMs exhibits more consistent line-by-line perplexity
compared to human-authored code~\citep{xu2024detecting}.
Still,
limited research has been conducted on code perplexity through LLMs.
To fill this gap
in this study we investigate code perplexity
across diverse programming languages, models, and datasets
using multiple LLMs.
Specifically,
we set out to answer the following research questions:

\begin{description}
  \item[RQ1] \emph{How does code perplexity vary across programming languages?}
  \item[RQ2] \emph{How do intrinsic language properties and code structure features affect code perplexity?}
  \item[RQ3] \emph{How does code perplexity vary across LLMs on a multilingual code corpus?}
  \item[RQ4] \emph{How do different evaluation datasets affect an LLM's measured code perplexity?}
\end{description}

RQ1 and RQ2 are broader in scope, as they establish the base
per-language and per-property perplexity patterns; RQ3 and RQ4 are
narrower by design, each isolating one axis of generalization (model
choice in RQ3, evaluation corpus in RQ4) while holding the other
fixed.

This study makes the following contributions:

\begin{itemize}
\item A reproducible method and an open-source toolkit
for computing code perplexity using multiple LLMs,
enabling standardized evaluations across programming languages and datasets.
\item The first large-scale, language-specific analysis of
code perplexity measured through LLMs,
offering new insights into how model confidence varies across programming languages.
\item Analysis of the relationship between language properties and code structure features
and LLM perplexity,
showing that strongly-typed and more recent languages generally exhibit lower perplexity,
while comments increase it.
\item An assessment regarding the generalizability of code perplexity across LLMs and benchmark datasets,
providing patterns that can inform model selection and training strategies.
\item Contextualization of the findings in relation to prior work
on LLM confidence, uncertainty, and correctness,
highlighting the practical relevance of perplexity as a lightweight diagnostic indicator
in code generation tasks.
\end{itemize}

In the following section we present an overview of related work
on code completion powered by LLMs, and
approaches for their evaluation.
We then describe the study methods in Section~\ref{sec:methods},
present the research results in Section~\ref{sec:results}, and
discuss their implications in Section~\ref{sec:discussion}.
The study is complemented by considering its limitations
in Section~\ref{sec:limitations},
followed by our concluding remarks in Section~\ref{sec:conclusion}.
We provide any references to source code through permalinks
based on Software Hash Identifiers (SWHIDs---\citealp{ISO18670:2025}).
According to published guidelines~\citep{ince2012case},
the code and produced data of our study are publicly available
online,\footnote{\url{https://doi.org/10.5281/zenodo.10886458}} and
can be used for replication or further empirical research.

\section{Related Work}
\label{sec:related}

We classify related work into three main axes:
i) code attributes and naturalness;
ii) code completion with LLMs; and
iii) evaluation of code LLMs.
We then present the novelty our research offers
with respect to existing research.

\subsection{Code Attributes and Naturalness}

The naturalness hypothesis~\citep{allamanis2018naturalness} suggests that code,
like natural language,
serves as a means of human communication~\citep{knuth1984literate}, and
shares exploitable linguistic properties.
Hindle \etal~\citeyearpar{hindle2012naturalness} showed that code,
modeled with \emph{n}-grams, is even more predictable than text.
Ray \etal~\citeyearpar{ray2016buggy} found buggy code to be more entropic, and
thus less natural to language models.
Various strategies have been developed to handle the high rate of new or rare vocabulary
introduced by identifier names~\citep{karampatsis2020vocab,sharma2024survey}.
Ultimately, code's repetitive nature,
well-captured by Natural Language Processing-driven models,
enables the development of effective tools for software engineers.

\subsection{Code Completion with LLMs}

Most studies on LLM-based code completion either propose new models or
conduct empirical research.
Notable open code LLMs include
\emph{StarCoder}~\citep{li2023starcoder},
\emph{CodeGeeX}~\citep{zheng2023codegeex},
\emph{LLaMA}~\citep{touvron2023llama, touvron2023llama2},
\emph{CodeLlama}~\citep{roziere2024code},
\emph{SantaCoder}~\citep{allal2023santacoder},
\emph{PolyCoder}~\citep{xu2022polycoder},
\emph{InCoder}~\citep{fried2023incoder}, and
\emph{CodeGen}~\citep{nijkamp2023codegen}.

Empirical research largely focuses on
\emph{Copilot}~\citep{wang2023review}, examining its
applications~\citep{moroz2022potential}, impact on
productivity~\citep{ZKLR24}, and developer
interaction~\citep{barke2022grounded}. \emph{Copilot} serves as both
an acceleration and exploration tool, enhancing productivity and
assisting with unfamiliar tasks~\citep{barke2022grounded}. An early
study showed that, despite its benefits, improvements are needed in
security, reliability, and effectiveness~\citep{moroz2022potential}.
Since then, and with the accompanying improvements in LLMs, use of
Copilot has been shown to increase developer productivity as the acceptance
rates of its suggestions correlate with productivity
gain~\citep{ZKLR24,cui:2024,shihab:2025,bakal:2025}.

\subsection{Evaluation of Code LLMs}

Here we review the main evaluation frameworks, surveys, and empirical studies on code LLMs.

\subsubsection{Evaluation Frameworks}

Liu \etal~\citeyearpar{liu2024evalplus} propose \emph{EvalPlus},
a benchmarking framework for assessing the functional correctness of LLM-generated code.
It expands an initial dataset using LLM- and mutation-based input generators
to validate completions.
Testing 14 LLMs including \emph{GPT-4} and \emph{ChatGPT} reveals
many previously undetected errors,
lowering \emph{pass@k} by 15.1\% on average.
This suggests existing evaluations may overestimate code correctness.

Ding \etal~\citeyearpar{ding2023static} propose a static evaluation framework
for Python code completions, parsing snippets into Abstract Syntax Trees (ASTs) and
analyzing them with \code{Pyflakes}.\footnote{\url{https://github.com/PyCQA/pyflakes}}
They evaluate 100K function completions from the \emph{CodeGen} models~\citep{nijkamp2023codegen},
finding most AST errors stem from incomplete generations due to length limits,
while static errors mainly involve \emph{Undefined Name} and \emph{Unused Variable}.
Higher temperatures degrade performance,
and larger model sizes show mixed effects---reducing undefined variables
but increasing undefined methods.
Errors in prompts often propagate to completions,
highlighting LLMs' in-context learning.

Wang \etal~\citeyearpar{wang2023recode} introduce \emph{ReCode}, a
robustness evaluation framework for LLM-generated code. It applies
natural, semantics-preserving transformations (\eg on docstrings,
variable names, code syntax and format). Evaluating models like
\emph{CodeGen}~\citep{nijkamp2023codegen},
\emph{InCoder}~\citep{fried2023incoder}, and
\emph{GPT-J}~\citep{wang2021gptj} shows that while larger models and
diverse pretraining data improve robustness, they reduce
generalizability. Models are particularly sensitive to syntax changes.

\subsubsection{Surveys}

Chang \etal~\citeyearpar{chang2024survey} review LLM performance across domains
including code generation.
They find that \emph{CodeGen-16B}~\citep{nijkamp2023codegen} performs comparably
to \emph{ChatGPT} with more parameters.
While \emph{ChatGPT} struggles with data structures and graph theory,
it surpasses college students in dynamic programming, greedy algorithms, and search.
\emph{GPT-4} excels in code generation, understanding, and reasoning.

Zheng \etal~\citeyearpar{zheng2024survey} review the evolution,
benchmarking, and future trends of code LLMs. The \emph{CodeLlama}
series~\citep{roziere2024code} leads in code generation, while
\emph{ToolCoder}~\citep{zhang2023toolcoder} excels in API-based code
generation. \emph{GPT-4} and \emph{GPT-3.5} dominate test case
generation. Comparisons show that, with similar parameter sizes, code
LLMs outperform general LLMs, and current State-of-the-Art (SOTA) code
LLMs surpass general LLMs in code generation.

Husein \etal~\citeyearpar{husein2025llms} review LLMs for code completion,
noting productivity gains across all granularity levels in IDEs.
Larger datasets help prevent overfitting,
but performance varies by language and dataset.
Dynamically typed languages like Python pose challenges due to their flexible type information,
while statically typed languages like Java provide clearer context.
In Python, the low relevance of context lines increases ambiguity,
making precise completions harder.

\subsubsection{Empirical Studies}

Izadi \etal~\citeyearpar{izadi2024language} evaluated three public code LLMs
(\emph{InCoder}---\citealp{fried2023incoder},
\emph{UniXcoder}---\citealp{guo2022unixcoder},
\emph{CodeGPT}---\citealp{lu2021codexglue})
using real auto-completion data.
The models perform better in common languages like Python and Java
but struggle with less popular ones such as Rust and Scala.
\emph{InCoder} is the most performant,
suggesting superior training data and objectives.
However, all models show a high failure rate (66.3\%),
with issues including incorrect predictions and poor semantics.
The authors recommend improvements in training, error tolerance, usability, and hybrid systems.

Another systematic evaluation compares SOTA code LLMs
(\emph{Codex}---\citealp{chen2021codex},
\emph{GPT-J~6b}---\citealp{wang2021gptj},
\emph{GPT-Neo~2.7b}---\citealp{black2021gptneo},
\emph{GPT-NeoX~20b}---\citealp{black2022gptneox},
\emph{CodeParrot}---\citealp{tunstall2022natural}) across programming
languages using perplexity~(\citealp{xu2022polycoder},
cf.~Section~\ref{ssec:research-novelty} below). Despite \emph{Codex}'s
commercial advantage, open-source models perform comparably in some
languages. To address the need for a multilingual code model,
\emph{PolyCoder} (2.7b parameters, GPT-2 based) was trained on 249GB
of code in 12 languages. It ranked fourth overall, behind \emph{Codex}
and \emph{GPT-Neo/J}, but led in C language performance.

Yang \etal~\citeyearpar{yang2024unveiling} empirically analyze
the memorization ability of large code models
using variations of the perplexity metric.
Lower perplexity often indicates higher memorization,
especially on seen examples.
Models memorize various content types including
documentation,
licenses,
sensitive information, and
code logic.
Memorization is influenced by factors such as
model size,
top-\emph{k} sampling,
output length,
number of outputs, and
fragment frequency,
suggesting that deduplication of training data can help reduce memorization.

Zhou \etal~\citeyearpar{lexin2024largerlessreliable} show that
larger, more instructable LLMs can be less reliable than smaller ones,
often producing incorrect answers to difficult questions missed by human evaluators.
Their study assesses several LLMs
using six reliability indicators focused on
response proportion,
prompting stability, and
difficulty concordance.

Liu \etal~\citeyearpar{liu2024empirical} present a case study
on using \emph{GPT-4} for safety-critical software generation,
introducing \emph{Prompt-FDC},
a prompt engineering method combining
functional requirements,
domain features, and
constraints.
Their approach enhances
code completeness,
compliance,
readability,
maintainability, and
comment coverage,
showing that with proper prompting,
LLMs can meet production engineering standards.

\subsection{Research Novelty}%
\label{ssec:research-novelty}

\paragraph{What is, and is not, novel}%

Perplexity is not a new concept for LLM evaluation; several
code-oriented studies including
\emph{PolyCoder}~\citep{xu2022polycoder} already report multilingual
or multi-model figures. The novelty of this work is not a single
definition, but the combination of: a confidence- and
comparability-driven design on a copyleft, LLaMA-complementary GitHub
sample; a reproducible strided sliding-window perplexity
implementation (Section~\ref{sec:ppl-implementation}) with reruns
for 18~checkpoints; and a broad set of empirical RQs
(properties of languages, structure features, cross-LLM agreement, and
cross-benchmark orderings) tied to the same file population.

\paragraph{Positioning with respect to PolyCoder}%

\emph{PolyCoder} introduces a new code LLM and situates it against
competitors~\citep{xu2022polycoder}, whereas we treat perplexity as an
instrument for evaluating model confidence in code completion under a
fixed, balanced GPL-centered file sample
(Section~\ref{sec:project-selection}). \emph{PolyCoder}'s evaluation
sets vary strongly in files per language; our preprocessed draw keeps
languages more evenly represented for per-language medians, which
matters when ranking languages. Their reported pipeline aggregates
disjoint chunks, summing chunk log-likelihoods; we implement the
strided sliding-window approach so that each scored token, except for
short file heads, is predicted with up to the full allocated context.
Moreover, we shorten the active window to the file length when a file
has fewer than 2048 tokens, so context is file-adaptive rather than a
small fixed cap on every file (Section~\ref{sec:ppl-implementation}).
\emph{PolyCoder} also uses a long (2048-token) window, but their
benchmark files are relatively short in the median (598 tokens), which
limits how many tokens receive a full long-context read in practice;
our balanced GPL sample, together with the stride, yields dense
coverage of each file under the strided policy.

\paragraph{Empirical distinctions beyond PolyCoder}%

Beyond dataset and model choices, our findings that extend
\emph{PolyCoder}'s multilingual perplexity evaluation are:
a systematic view of how 18 public
checkpoints agree on language orderings, with a correlation map and
parallel coordinates (Section~\ref{sec:rq3}); a quantified test of
whether median language orderings carry over to the \emph{PolyCoder}
file set under the same model (Section~\ref{sec:rq4}); probes for
comments, typing, age, and size, including tests where appropriate
(Section~\ref{sec:rq2}), along with a discussion that separates how
much model change versus benchmark change moves perplexity, which was
not the target of the original \emph{PolyCoder} study; and a
leakage-aware GPL sampling rationale tied to the training data of
specific evaluator models (\ie controlling known overlap with LLaMA's
and StarCoder~2's training data),
which is orthogonal to merely listing per-language perplexities in a
multi-language table. When we agree directionally with prior
observations (\eg the role of static type
hints---\citealp{husein2025llms}), our contribution is complete quantitative
results on this corpus, a cross-model and cross-benchmark validation,
and a clear discussion of limitations from broad language categories and
limited sample size ($n$).

\section{Methods}
\label{sec:methods}

Our methods involve
selecting and preprocessing the code dataset of our analysis as well as
selecting, configuring, and running an LLM for computing the perplexity of the dataset.
An overview of the methods is presented
through a UML information flow diagram in Figure~\ref{fig:methods}.

\begin{figure}[t]
  \centering

\scalebox{0.75}{
\begin{tikzpicture}[>=Stealth,
  box/.style={draw, align=center, inner ysep=1ex},
  datasrc/.style={box, fill=teal!20},
  projprep/.style={box, fill=violet!20},
  fileprep/.style={box, fill=orange!20},
  llmeval/.style={box, fill=blue!20},
]

\begin{scope}[start chain=going below,
  every node/.style=box,
  node distance=.5cm,
  every on chain/.style={join={by ->, dashed}}
  ]
  
  \node[on chain, datasrc] at (0, 0) (0) {Google
    BigQuery\\GitHub dataset}; 
  
  \node[on chain, projprep] (1) {Project license filtering};

  \node[on chain, projprep] (2) {Project deduplication};

  \node[on chain, projprep] (3) {Project quality filtering};

  \node[on chain, projprep] (4) {Project sampling};

  \node[on chain, projprep] (5) {Project bare cloning};

  \node[on chain, fileprep] (6) {File quality filtering};

  \node[on chain, fileprep] (7) {File deduplication};

  \node[on chain, fileprep] (8) {File sample selection};

  \node[on chain, fileprep] (9) {Code cleaning};

  \node[on chain, llmeval] (10) {Perplexity calculation};

  \node[on chain, llmeval] (11) {Evaluation of code\\perplexity through LLMs};
\end{scope}

\begin{scope}[xshift=2cm]
\node[draw, dashed, align=left, inner sep=6pt, anchor=north west] at (2.0, 0.5) {%
  \colorbox{teal!20}{\hspace{0.45cm}\vphantom{X}}~Data source\\[3pt]
  \colorbox{violet!20}{\hspace{0.45cm}\vphantom{X}}~Project preprocessing\\[3pt]
  \colorbox{orange!20}{\hspace{0.45cm}\vphantom{X}}~File preprocessing\\[3pt]
  \colorbox{blue!20}{\hspace{0.45cm}\vphantom{X}}~LLM evaluation\\[3pt]
  \tikz[baseline=-0.5ex]\draw[dashed,->](0,0)--(0.65cm,0);~«flow»%
};
\end{scope}

\end{tikzpicture}
}
  \caption{Information flow of study methods.}
  \label{fig:methods}
\end{figure}

\subsection{LLM Selection}
\label{sec:llm-selection}

We used a set of general-purpose and code LLMs supported by
\code{llama.cpp}:\footnote{%
\url{https://github.com/ggml-org/llama.cpp}
}
\mbox{CodeLlama~7b~and~13b}~\citep{roziere2024code};
\mbox{CodeShell~7b}~\citep{xie2024codeshelltechnicalreport};
\mbox{Gemma~3~27b}~\citep{gemmateam2025gemma3technicalreport};
\mbox{Gemma~4~31b}~\citep{farabet2026gemma4};
\mbox{LLaMA~2~7b~and~13b}~\citep{touvron2023llama2};
\mbox{LLaMA~3~8b}~\citep{dubey2024llama3herdmodels};
\mbox{LLaMA~3.1~8b}~\citep{dubey2024llama3herdmodels};
\mbox{LLaMA~3.2~1b~and~3b}~\citep{meta2024llama3_2};\footnote{%
  Within the LLaMA~3.2 release, Meta ships
  lightweight \emph{text-only} checkpoints at 1b and 3b parameters,
  whereas the larger 11b and 90b LLaMA~3.2 variants are multimodal
  vision-language models~\citep{meta2024llama3_2}. Because perplexity
  here is computed over textual code tokens, we restrict attention to
  these text-only LLaMA~3.2 weights rather than the vision-capable
  checkpoints.}
\mbox{Mistral~7b}~\citep{jiang2023mistral7b};
\mbox{Mixtral~MoE~8x7b}~\citep{jiang2024mixtralexperts};
\mbox{Qwen~3.5~9b~and~27b}~\citep{qwen35blog}; and
\mbox{StarCoder~2~3b,~7b~and~15b}~\citep{lozhkov2024starcoder2stackv2}.
The checkpoints were retrieved from Hugging Face and
converted to \code{GGUF F16} format
with the \code{convert\_hf\_to\_gguf.py} script
by \code{llama.cpp}.\footnote{%
  \url{https://archive.softwareheritage.org/swh:1:cnt:42d559dfecf91ed33dfc9183f586eef3c6a2b9f9;origin=https://github.com/ggml-org/llama.cpp;visit=swh:1:snp:481de56e1a7d5e2e6a03219a657d2bbf677a0f5d;anchor=swh:1:rev:45cac7ca703fb9085eae62b9121fca01d20177f6;path=/convert_hf_to_gguf.py}
}
\mbox{Mixtral~MoE~8x7b} was additionally quantized to \code{Q4\_K\_M}
format with \code{llama.cpp}'s \code{llama-quantize} tool to fit into
the GPU memory of our laboratory machine.

\subsection{Project Selection}
\label{sec:project-selection}

To avoid evaluation overfitting we synthesized the dataset of the
analysis with code that is complementary to the code used to train
LLaMA~\citep{touvron2023llama}. Specifically, we created a dataset of
GitHub projects that were \emph{not} used in LLaMA's training dataset,
by selecting projects distributed under (copyleft) GNU licenses, as
detailed below. LLaMA is the only model whose training-data
documentation identifies the licenses of a component of its training
data in enough detail to support this exclusion. StarCoder~2 is
the one model in our study of 18~models (Section~\ref{sec:llm-selection})
with its own documented license-based filter: its \emph{The
Stack~v2}~\citep{lozhkov2024starcoder2stackv2} training corpus reports
excluding copyleft-licensed files, so restricting our sample to
GNU-licensed projects also avoids known overlap with its training
data. For the remaining models we cannot confirm whether GPL-licensed
code was included in or excluded from their training; we discuss the
scope of this exclusion further in Section~\ref{sec:limitations}.

In LLaMA the authors queried the public \emph{GitHub Activity Data}
available on Google
BigQuery,\footnote{\url{https://console.cloud.google.com/marketplace/product/github/github-repos}}
and only kept projects that are distributed under the (permissive)
Apache, BSD, and MIT licenses. Consequently, we used the same dataset
(snapshot 23 Feb 2023), and only kept (copyleft) GNU-licensed
projects. For this we queried Table \emph{licenses} (snapshot 27 Nov
2022), and
identified 15 distinct licenses in all projects.
From these we filtered all copyleft licenses
(\emph{gpl-2.0, gpl-3.0, agpl-3.0, lgpl-2.1, lgpl-3.0}).
We then selected from the table \emph{licenses} all projects distributed with these licenses.
The total number of retrieved projects was 785\,567.

\subsection{Project Preprocessing}
\label{sec:project-preprocessing}

\paragraph{Project deduplication}%

We deduplicated the GPL projects on the project level using the
deduplication dataset by Spinellis
\etal~\citeyearpar{spinellis2020deduplication}. We mapped projects to
the source field of the deduplication dataset, and used the
corresponding target project. We kept projects not present in the
deduplication dataset, and also verified that final projects are not
present in the LLaMA initial project list.
From this process we kept 757\,642 unique projects.

\paragraph{Project quality filtering}%

We filtered projects based on the following recommended quality
criteria~\citep{dabic2021sampling, kalliamvakou2015mining,
  sheoran2014watchers}. We kept projects with at least one star, one
fork, and a listed primary language. In addition, we filtered out
projects where the frequency of the primary language within the
projects was less than 1\%. To retrieve these project metadata we
queried the GitHub GraphQL
API.\footnote{\url{https://docs.github.com/en/graphql}}
In the end a total of 114\,160 projects were retained.


\paragraph{Project sample selection}%

We selected a random sample of 1008 projects for experimentation
uniformly from all 14 languages included in the dataset (\ie 72
projects per language),\footnote{ C, C\#, C++, CSS, Go, HTML, Java,
  JavaScript, Perl, PHP, Python, R, Ruby, Shell} and used Git to clone
the \emph{bare} projects with $\mathit{depth}=1$. We strived for a
sample size that was large enough to provide a representative dataset,
but also manageable by our laboratory infrastructure and time
constraints. The employed \emph{git-clone} configuration minimizes
storage wasted on data not needed by our study.

\subsection{File Preprocessing}
\label{sec:file-preprocessing}

\paragraph{File quality filtering}%

We kept 430\,359 files whose extensions belonged to the 14 languages
of the dataset. We created a list of file extensions per language
based on the list of
extensions\footnote{\url{https://archive.softwareheritage.org/swh:1:cnt:af779c91e6a52ae9ffbd8fb566ae50e461d066f5;origin=https://github.com/github-linguist/linguist;visit=swh:1:snp:26c84095ee29473e8ec4ffa6bebc095d7ae47bf9;anchor=swh:1:rev:af34cb5e7ba4a40721f65a782e0e8ddacfbfe459;path=/lib/linguist/languages.yml}}
provided by the \code{github-linguist/linguist}
project.\footnote{\url{https://github.com/github-linguist}} We
converted the extensions list to CSV format, and manually reviewed the
extensions of the languages with $\mathit{frequency} \geq 1\%$ to
remove any redundant extensions.

\paragraph{File deduplication}%

Similarly to LLaMA training,
we deduplicated exact-matched files based on the Git-provided SHA sums,
retaining 296\,543 unique files.
From an analysis of the SHA sums of the sampled projects it appears that
duplicate files are mostly empty \code{\_\_init\_\_.py} and empty-body \code{index.html} files
(\code{<html><body>\allowbreak</body></html>}).



\paragraph{File sample selection}%

We classified each sampled project into multiple languages based on the extensions of its files.
We then performed language-level file sampling while ensuring a
balanced representation of the projects using a combination of
stratified and random sampling. For sample size we used the number of
files of the language with the lowest file representation (\ie
R-Project, 161 files). We did this to include as many files as
possible but also equal in number across languages. Therefore, for
each language, we sampled 161 files in total from as many projects as
possible,
leading to a final set of 2254 sampled files from 881 distinct projects.


\paragraph{Code cleaning}%

We first decoded all 2254 source code files after detecting their
encoding with the \code{chardet} Python
library.\footnote{\url{https://github.com/chardet/chardet}} Next we
removed any boilerplate header comments from the files, \ie
non-code-related headers such as license notices, copyright
statements, version tags, and author attributions. For this we used
the \code{pygments} library\footnote{\url{https://pygments.org/}} to
tokenize the files, and then detected and removed the following
comment types occurring at the file start (until the first non-comment
line is found): \emph{Comment.Multiline}, \emph{Comment.Single}, and
\emph{Comment.Special}. We skipped any comments of type
\emph{Comment.Hashbang} or \emph{Whitespace}. We also kept any
\emph{String.Doc} header comments because these mainly include
code-related comments. To conclude that we identified all
\emph{String.Doc} header comments in our sampled projects, and
manually reviewed a random sample of 50 such comments.

\subsection{Perplexity Implementation}%
\label{sec:ppl-implementation}

Mathematically, perplexity is defined as the exponentiated
cross-entropy loss of a prediction sequence~\citep{jelinek:1977}.
Given a tokenized sequence $X=(x_1, x_2, ..., x_N)$ it can be
calculated as:

\begin{equation}
\mathrm{Perplexity}(X) =
2^{-\frac{1}{N}\sum_{i=1}^{N}\lg{p\left(x_i|x_{<i}\right)}}
\end{equation}

In LLMs, the tokenized sequence is the sequence of predictions. If the
model is very confident in its prediction, each $p(x_i|x_{<i})$ will
be equal to one and the perplexity will be one. If the model gives
equal probability to each word for all predictions, then the
perplexity will be equal to the size of the vocabulary.

Initially we experimented with Hugging Face's method for computing
perplexity for fixed-length
models.\footnote{\url{https://web.archive.org/web/20250518012847/https://huggingface.co/docs/transformers/perplexity}}
The method comes with an implementation example with GPT-2 using the
Hugging Face Transformers
library.\footnote{\url{https://web.archive.org/web/20250511140609/https://huggingface.co/docs/transformers/index}}
We used this implementation after replacing the \emph{WikiText-2}
dataset with our preprocessed (untokenized) source code files, and
GPT-2 with one of our evaluator models,
LLaMA~3.2~3b.\footnote{\url{https://web.archive.org/web/20250419160227/https://huggingface.co/blog/llama32}}
However, this implementation was too slow on our laboratory machines,
and we also detected some potential memory leakage.

Therefore, we developed a Python driver for file-level perplexity
that loads a \code{llama.cpp} model
via the Python bindings provided by the \code{llama-cpp-python} library.\footnote{%
\url{https://github.com/abetlen/llama-cpp-python}
}
We used the \code{cuBLAS} build, which provides BLAS acceleration
using the CUDA cores of our NVIDIA A100 PCIe 80 GB GPU\@.
This stack was fast enough for our large-scale runs, and
did not exhibit the slowdown and suspected memory issues we saw
with the Transformers-based prototype.

Although perplexity is mathematically defined as above, it is not
feasible, nor optimal, to compute it following the exact mathematical
formula. Regarding optimality, LLMs generate tokens using previous
tokens as context. Therefore to guard against volatile predictions
because of lack of sufficient context, we want to calculate perplexity
after an initial number of tokens in the sequence. Regarding
feasibility, LLMs have limits on the number of tokens they can take as
input (their context window). That means that we have to break the
input into windows and feed them to the LLM\@. Even when LLMs have
long-enough windows (and they get longer with time), using them
requires more resources, which may be beyond a study's limits.

There are two ways to break the sequence into windows that constitute
the context. In the first approach, windows are not overlapping. All
tokens in the context are evaluated by causal attention. Then we start
at the middle of the context window and calculate the negative
log-likelihood of the prediction of the first, second, \etc, tokens
from that point until the end of the window. Therefore the first token
from the middle uses half of the context for its perplexity, the
second token from the middle adds the first token in the context of
its perplexity, until the last token of the window that has all the
previous tokens of the window as its context. Note that as the windows
are non-overlapping, only half of the tokens of the sequence are used
for calculating the perplexity; see
Figure~\ref{fig:perplexity-non-strided}.

In the second approach, the windows are overlapping; the length of
their non-overlapping part is the \emph{stride}. Perplexity is
calculated over the last stride tokens of each sliding window. That
means that only the tokens of the first window up to the stride do
not have their perplexity calculated; see
Figure~\ref{fig:perplexity-strided}. For example, given a sequence of
2000 tokens and a context size of 1024, the sliding window strategy
with stride equal to 1 produces overlapping chunks such as tokens 0--1023,
1--1024, 2--1025, and so on. Each window is used to evaluate the
log-likelihood of its final token.

\begin{figure}[t]
  \centering
  \begin{subfigure}[t]{0.48\textwidth}
    \centering
    \resizebox{\linewidth}{!}{%
      \begin{tikzpicture}
        \def\barh{0.8}
        \foreach \step/\lbl in {0,...,3} {
          \pgfmathsetmacro{\x}{3*\step}
          \pgfmathsetmacro{\y}{-1.2*\step - 1.2}
          \draw (0,\y) rectangle (12, \y+\barh);
          \draw[pattern={north east lines}] (\x,\y) rectangle (\x+3,\y+\barh);
          \draw[pattern={north west lines}] (\x+1.5,\y) rectangle (\x+3,\y+\barh);
        }
      \end{tikzpicture}%
    }
    \caption{}
    \label{fig:perplexity-non-strided}
  \end{subfigure}\hfill
  \begin{subfigure}[t]{0.48\textwidth}
    \centering
    \resizebox{\linewidth}{!}{%
      \begin{tikzpicture}
        \def\barh{0.8}
        \foreach \step in {0,...,3} {
          \pgfmathsetmacro{\y}{-1.2*\step - 1.2}
          \draw (0,\y) rectangle (12, \y+\barh);
          \draw[pattern={north west lines}] (\step+2,\y) rectangle (\step+3,\y+\barh);
          \draw[pattern={north east lines}] (\step,\y) rectangle (\step+3,\y+\barh);
        }
      \end{tikzpicture}%
    }
    \caption{}
    \label{fig:perplexity-strided}
  \end{subfigure}
  \caption{
    Perplexity calculation with sliding window context and:
    (a) without stride,
    (b) with stride.
    In both panels, context window is hatched, perplexity calculation is on the
    cross-hatched regions.
  }
  \label{fig:perplexity-window-schemes}
\end{figure}

The strided method is more accurate, but requires more calculations:
the computation for stride equal to~1 is context size times more than
the non-stride calculation. As a compromise, our perplexity program
loads each model with context size set to~2048, and invokes the driver
with stride equal to~512; to ensure adequate context, it uses the
first~512 tokens of the first window as context for starting the
predictions with the following token. When a file has fewer than 2048
tokens the implementation shortens the working window to the tokenized
file length and the starting context to half that, so the effective
context is adaptive per file instead of padding or forcing a very
small static cap. All figures in Section~\ref{sec:results} that
reference this configuration use the same strided policy; RQ4 reuses
it on the \emph{PolyCoder} files.

\section{Results}%
\label{sec:results}

Below we present our results with respect to
the research questions outlined in Section~\ref{sec:intro}.
RQ1 and RQ2 establish \emph{what} the per-language and per-property
perplexity patterns are, aggregated across all 18~models
(Section~\ref{sec:llm-selection}); RQ3 and RQ4 then test \emph{how
consistent} those patterns are when we vary the model (RQ3, within our
corpus) or the evaluation corpus (RQ4, across the same 18~models).
Our key takeaways are set in \emph{italics} in shaded boxes.

\subsection{RQ1: How does code perplexity vary across
  programming languages?}%
\label{sec:rq1}

\begin{figure}[t]
  \centering
  \includegraphics[width=\textwidth, keepaspectratio]{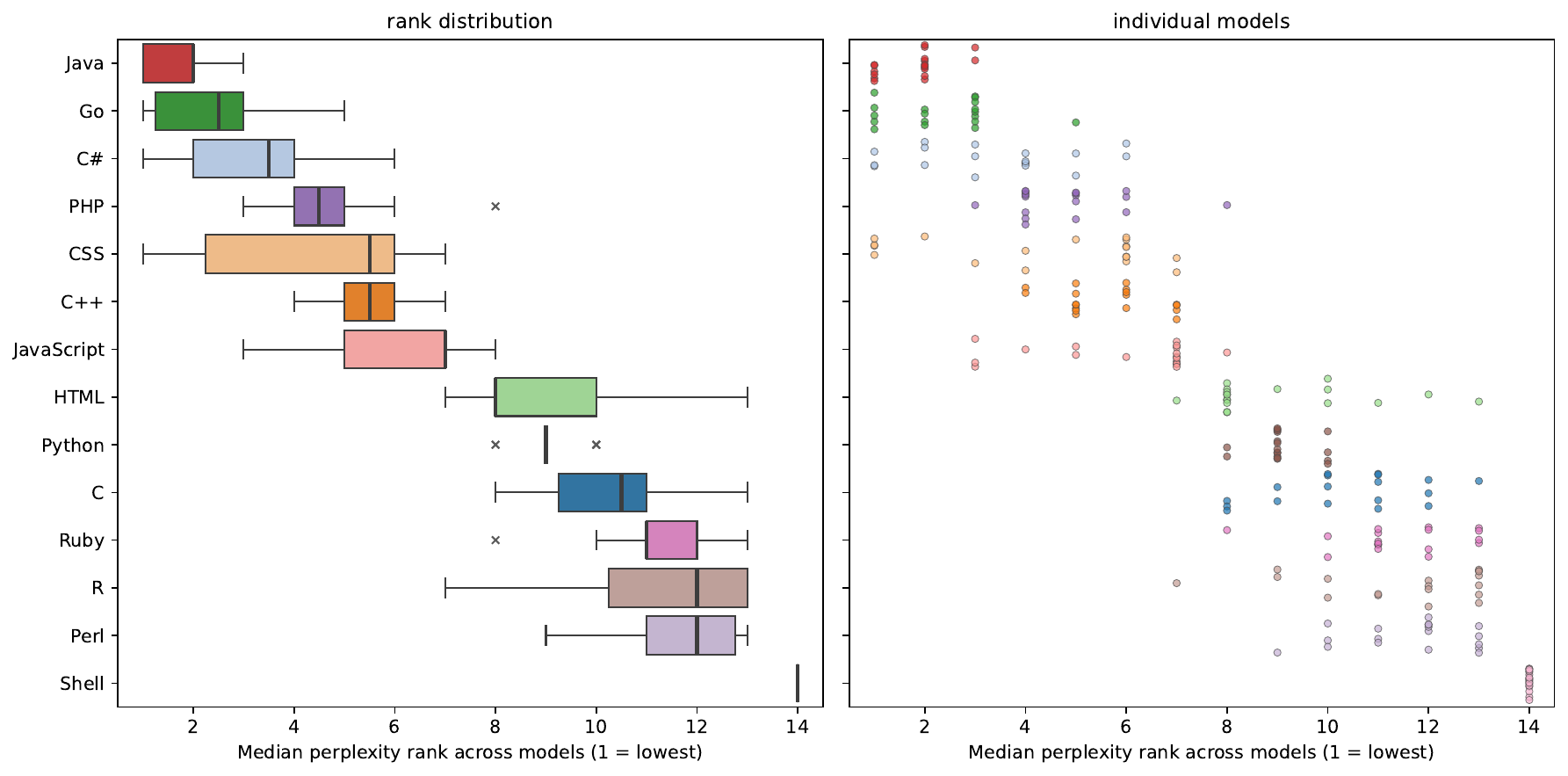}
  \caption{Perplexity across programming languages.}%
  \label{fig:perplexity-language-rank}
\end{figure}

In Figure~\ref{fig:perplexity-language-rank} we present perplexity box
plots and a strip plot per programming language for all the models we
used. To get around the fact that raw perplexity is not comparable
across models, we ranked languages by median perplexity within each
model in ascending order of median perplexity rank. Lower perplexity
suggests that the model has a better predictive ability regarding the
code syntax and semantics, leading to more accurate predictions. We
observe that Java, Go, C\#, and PHP have the lowest (best) perplexity,
while Shell, Perl, R, Ruby, and C demonstrate the highest (worst)
perplexity. Perl and R are tied for second place; the tie-breaker is
mean rank. Similarly, C++ and CSS are tied and the tie is resolved in
the same way.%
\footnote{
We combine per-model ranks across models in one of two ways, chosen by
what the result is for. Where we need an ordering of the languages
(the panels of Figure~\ref{fig:perplexity-language-rank}, and later on
the comparisons against other studies in Section~\ref{sec:rq3}, and
the corpus comparison in Section~\ref{sec:rq4}) we take the median of
each language's per-model rank, breaking ties by the mean as above.
The median is robust to a single model that agrees poorly with the
rest, and it is what the median line inside each box shows, so the
boxes never appear out of sequence. Where a rank instead serves as a
continuous variable to correlate against another quantity (the size,
vocabulary, and release-year comparisons of Section~\ref{sec:rq2}) we
take the mean, which over these 14 languages leaves no ties, whereas
the median ties four of them.}

We see two languages, Shell and Python, where the whiskers collapse at
the median, but each has its own story. For Python, 12~of the models
rank it at position~9, 2~at position~8, and~4 at position 10. Because
$12/18=67\%$ of the models are at exactly~9, both the 25th and the
75th percentiles are located there, so $\textrm{Q1} = \textrm{Q3} = 9$
and $\textrm{IQR} = 0$ and the whiskers collapse at the median. Shell,
however, has unanimously the highest perplexity for all models, so the
whiskers of its box plot collapse at the median because there is only a
unique distinct value, without variation. The strip plot at the right
complements the box plots as it allows us to differentiate between the
two languages.

Overall, code perplexity varies by programming language.
Developers can use this ranking to assess which of their software
projects, based on their language, could derive greater benefit from
LLM assistance. Furthermore, one can argue that less readable
languages have higher perplexity~\citep{husein2025llms}; Perl, for
instance, has been famously described as a ``write-only''
language~\citep{SFP11}.

\finding{%
  Code perplexity varies by programming language.
}

\begin{figure}[t]
  \centering
  \includegraphics[width=0.75\textwidth, keepaspectratio]{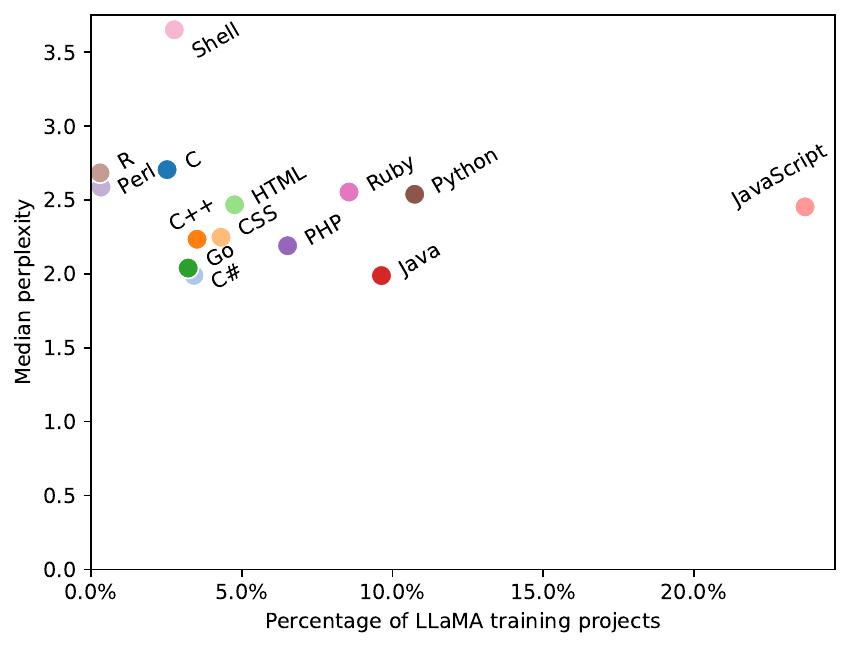}
  \caption{Language distribution in LLaMA training data versus perplexity.}%
  \label{fig:llama-language-distribution-perplexity}
\end{figure}

A possible factor for the differences in perplexity could be that
models have been trained on more projects in some languages and then
could be more confident in them. We checked this on LLaMA, by
investigating the effect of the language distribution in the LLaMA
training data on code perplexity. We retrieved the primary language of
the projects that we found to be in the LLaMA training data (see
Section~\ref{sec:project-selection}) by querying again the GitHub
GraphQL API as in Section~\ref{sec:project-preprocessing}.
Figure~\ref{fig:llama-language-distribution-perplexity} plots, for
each language, the percentage of projects in the LLaMA training data
against median code perplexity.

JavaScript has the largest training share, but its median perplexity
is only mid-range. Java combines high training share with low median
perplexity, while Shell combines low share with higher perplexity.
Because the sample contains only 14 languages, those contrasts can
still steer an informal fit slightly downward. That tilt reflects
leverage from a few languages rather than a robust pattern across all
of them. We quantified the association over the $n=14$ languages, but
we found no statistical significance at $\alpha=0.05$:
Pearson~$r=-0.14$ ($p = 0.62$) and Spearman~$\rho=-0.44$
($p = 0.12$). Taking out JavaScript and Shell as outliers did
not result in statistically significant results. Taken together, these
results do not support a systematic effect of training-language
distribution on code perplexity.

\finding{%
  The hypothesis that the language distribution in the LLaMA training
  data affects code perplexity is not supported. }

\subsection{RQ2: How do intrinsic language properties and code structure features affect code perplexity?}%
\label{sec:rq2}

After observing perplexity variation across programming languages we
aimed to identify which intrinsic language properties and code
structure features might explain these differences in model
performance. Specifically, we investigated the effect of code
comments, language typing, language age, code size and vocabulary size
on code perplexity.

To investigate the effect of comments on code perplexity we
preprocessed the sampled files of all languages by removing any
boilerplate header comments (see
Section~\ref{sec:file-preprocessing}). We proceeded to investigate
whether comments do affect perplexity by re-running the perplexity
program on a second version of the same sample of files, produced by
the preprocessing of Section~\ref{sec:file-preprocessing} with the code
cleaning step additionally removing all remaining comments (inline and
block comments throughout the file body). The two versions therefore
contain exactly the same files and share the same boilerplate-free
baseline; the only difference is the presence or absence of those
in-body comments, which is what allows the per-file pairing used below.

\begin{figure}[t]
  \centering
  \includegraphics[width=0.75\textwidth, keepaspectratio]{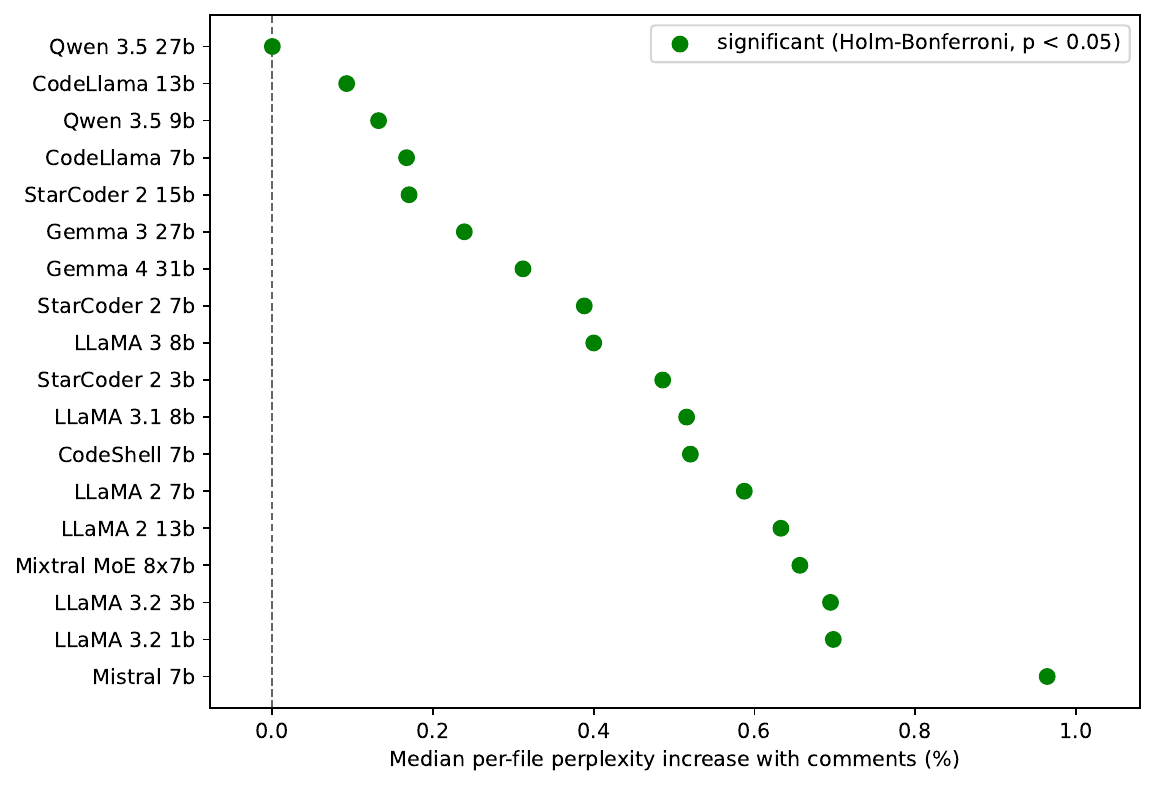}
  \caption{Median per-file relative perplexity increase attributable to
    comments, per model. Each point is the median over the 1516 paired
    files; the dashed line marks no change.}
  \label{fig:perplexity-comments-effect}
\end{figure}

At a file level, comments increase perplexity, albeit modestly. We
carried out a paired Wilcoxon test for each model, for 1516 pairs of
files that were scored and contained comments and not only comments.
We found that across all models perplexity rose (we applied the
Holm-Bonferroni correction for multiple tests); moreover, it rose in
50\% or more of the files, for each model. The median relative change
of perplexity ranged from 0\% (Qwen~3.5~27b) to 0.96\% (Mistral~7b);
see Figure~\ref{fig:perplexity-comments-effect}. A plausible
hypothesis is that natural language in comments is on average less
predictable, under the same next-token model, than the surrounding
program tokens. However, this does not affect the language perplexity
rankings. In Figure~\ref{fig:perplexity-comments-effect} every model's
point falls on the positive side of the zero reference line, although
for Qwen~3.5~27b it is barely distinguishable from it.\footnote{%
  Since the study's goal is not to evaluate prediction models or
  create a new prediction model, in the remainder of the analysis we
  retain the initial boilerplate-cleaned (but comment-inclusive) file
  sample.}

\begin{figure}[t]
  \centering
  \includegraphics[width=0.75\textwidth, keepaspectratio]{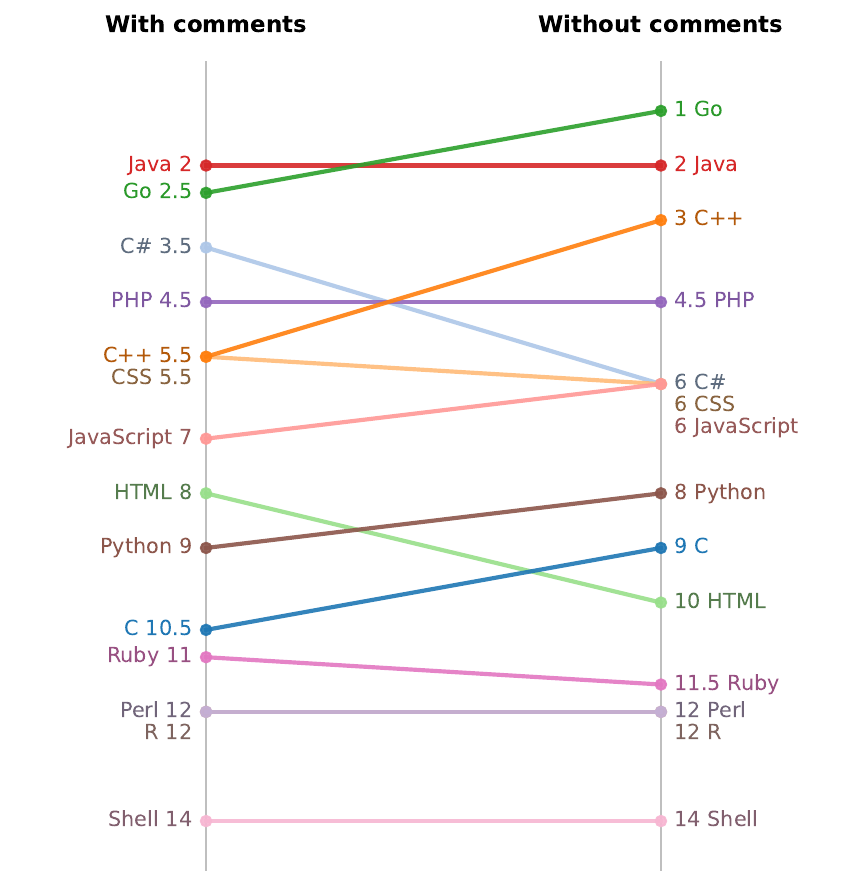}
  \caption{Per-language median perplexity rank across all models, with
    and without comments ($1$ = lowest perplexity).}
  \label{fig:perplexity-comments-rank}
\end{figure}

At the language level, we can investigate whether the language
rankings (as shown in Figure~\ref{fig:perplexity-language-rank}) are
affected when we exclude comments. In
Figure~\ref{fig:perplexity-comments-rank} we see that even though
languages do move (7/14 languages move by at least one position), the
overall ranking is not affected significantly. Indeed, the correlation
between the two language rankings (with and without comments) is very
high (Spearman $\rho = 0.95$ with $p \ll 1$, Kendall $\tau = 0.84$
with $p \ll 1$).

\finding{%
  Code comments increase perplexity modestly but do not affect the
  language perplexity ranking.}

By observing the language ranking of
Figure~\ref{fig:perplexity-language-rank} we identify a common
characteristic of the languages placed at the top and bottom of the
ranking: the top-ranked languages are mainly strongly typed, whereas
the bottom-ranked ones are mainly dynamically typed languages. A
special case here is Python, which can exhibit characteristics of both
strong and dynamic typing---which perhaps explains its place near the
middle.
In our dataset only one Python sample contains the arrow operator
(``\verb|->|'') in positions consistent with typed return annotations
under our simple token-based scan, so we treat Python here as leaning
toward dynamic usage in practice. Scripting-oriented
languages~\citep{scott2025pragmatics} tend to sit higher in the
ranking, which is interesting considering that they often have simpler
surface syntax than some of the top-ranked languages, which can be
more verbose.

These observations align with previous
research~\citep{husein2025llms}, but are based on~14 languages and a
coarse binary split. To complement the narrative, we divided languages
into a strongly typed set (C, C\#, C++, CSS, Go, Java) and a
scripting-oriented set (HTML, JavaScript, Perl, PHP, Python, R, Ruby,
Shell) and compared the two sets with a two-sided Mann-Whitney~$U$
test on the mean across models of each language's median perplexity
rank. We must note, however, that languages are neither independent
nor exchangeable units (for instance, we cannot meaningfully simply
swap C and Python between the two groups, as required by the
Mann-Whitney $U$ test), labels are debatable (\eg Python), and $n=14$
is small. We obtained $U=5$ and $p \approx 0.013$ (two-sided). The
probability that strongly typed languages have a better (lower) rank
is $0.896$ and the rank-biserial is $r_{rb} = 0.792$. Checking with
the per-language median ranks, to make sure the conclusion is not a
result of aggregating the per-model ranks with the mean, we obtained
similar results, as happened also with one-sided Mann-Whitney $U$
tests.

We then checked which languages buck the trend, \ie strongly typed
languages that rank worse than scripting languages. We found five such
instances: C ranked worse than HTML, JavaScript, PHP, and Python, and
C++ ranked worse than PHP\@.

Going down at the individual model level, we found that for all
18~models strongly typed languages had better (lower) median
perplexity than scripting languages, although the results were not
statistically significant per the Holm-Bonferroni correction apart
from a single model, CodeShell~7b. That is a result of the study
design itself, and in particular the number of languages; it could not
be fixed just by adding more models.

\finding{%
  Coarse grouping suggests lower median perplexity for strongly typed
  languages than for scripting-oriented languages in our sample.}

\begin{figure}[t]
  \centering
  \begin{subfigure}[t]{0.48\textwidth}
    \centering
    \includegraphics[width=\linewidth, keepaspectratio]{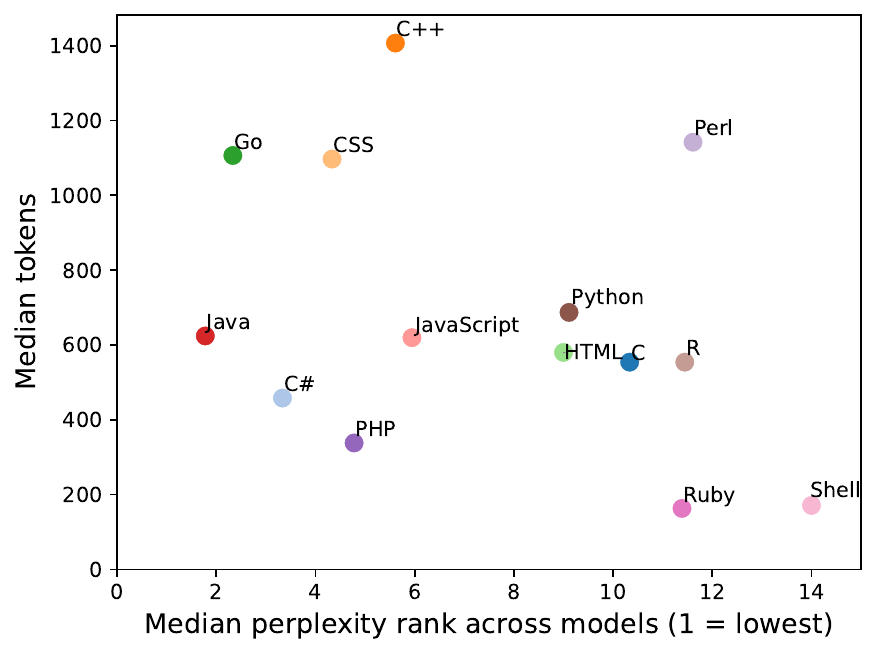}
    \caption{}
    \label{fig:size-perplexity-rank}
  \end{subfigure}\hfill
  \begin{subfigure}[t]{0.48\textwidth}
    \centering
    \includegraphics[width=\linewidth, keepaspectratio]{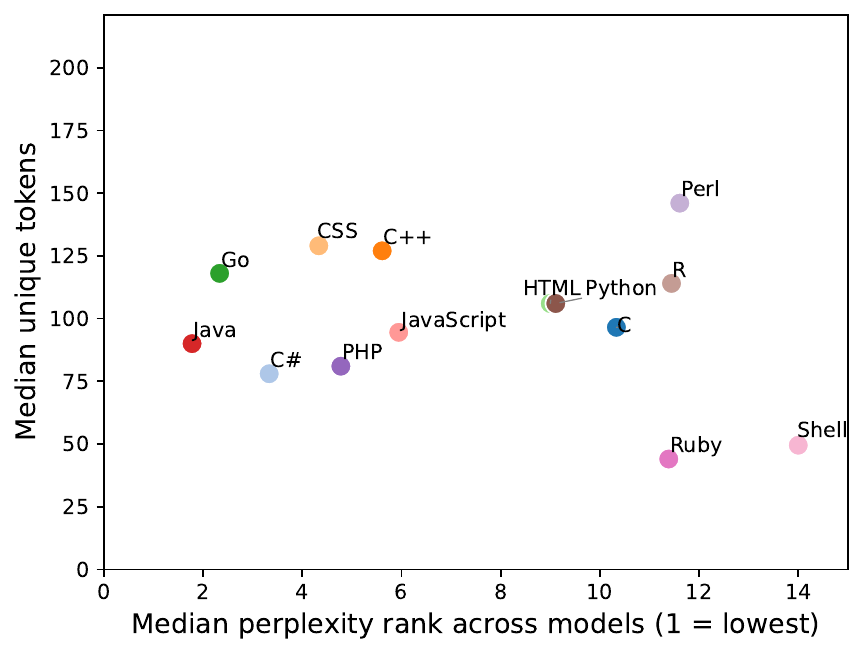}
    \caption{}
    \label{fig:voc-perplexity-rank}
  \end{subfigure}
  \caption{Per-language median total tokens (a) and median unique tokens (b)
    versus median perplexity rank across models.}
  \label{fig:size-voc-perplexity}
\end{figure}

We also investigated whether perplexity correlates with the total size
or the vocabulary size of source code, influenced by the work of
Karampatsis \etal~\citeyearpar{karampatsis2020vocab}, who demonstrate
that large vocabularies can negatively impact model performance and
scalability.\footnote{Perplexity, by definition, depends on the
  vocabulary size; but the dependency is on the vocabulary size of the
  LLM, not the vocabulary size of the programming language.}
Figures~\ref{fig:size-perplexity-rank}
and~\ref{fig:voc-perplexity-rank} plot, for each language, the number
of median total tokens and median unique tokens of the preprocessed
files against median perplexity rank. We computed Pearson and Spearman
correlations between those language-level medians, as we did for
Figure~\ref{fig:llama-language-distribution-perplexity} in
Section~\ref{sec:rq1}. For total tokens, Pearson~$r=-0.364$
($p = 0.200$) and Spearman~$\rho=-0.301$ ($p = 0.295$); for unique
tokens, Pearson~$r=-0.215$ ($p = 0.461$) and Spearman~$\rho=-0.044$
($p = 0.881$). We also checked the associations without Ruby and
Shell, which appear as outliers. This weakens the association for
total tokens, to $r=-0.066$ ($p = 0.840$) and $\rho=-0.011$
($p = 0.974$), and reverses the sign of the association for unique
tokens, to $r=0.378$ ($p = 0.226$) and $\rho=0.361$ ($p = 0.249$). In
fact, it reverses to $r=0.021$ ($p = 0.947$) even by excluding Shell
alone. That one of the 14 languages causes the sign to flip is a
sign in itself of how feebly the associations are supported. None of
these associations was statistically significant at $\alpha=0.05$; we
therefore do not infer an effect of file size or vocabulary size on
perplexity from these summaries.

\finding{%
  Language-level correlational tests do not support a reliable association
  between total size or vocabulary size of source code and perplexity.
}

\begin{figure}[t]
  \centering
  \includegraphics[width=0.75\textwidth, keepaspectratio]{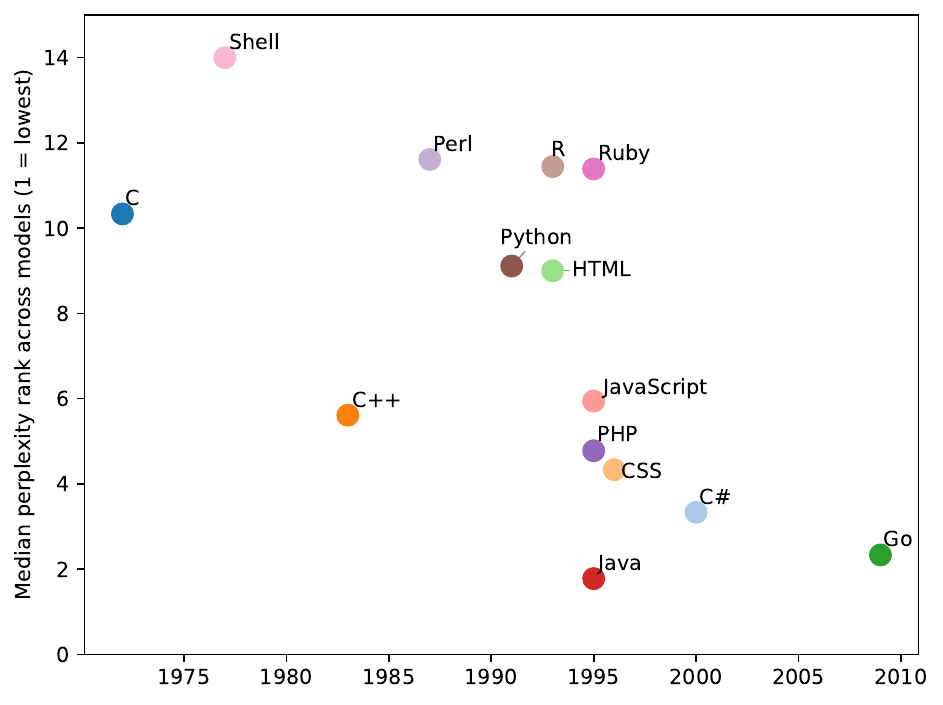}
  \caption{Release dates of programming languages versus median
    perplexity rank across models.}%
  \label{fig:year-perplexity-rank}
\end{figure}

Furthermore, we investigated how perplexity relates to each
programming language's age. This was motivated by the observation that
language design principles have evolved over time, with more recent
languages typically featuring stronger typing systems and more
standardized syntax---characteristics that Husein
\etal~\citeyearpar{husein2025llms} found to provide clearer context
for code prediction models. In Figure~\ref{fig:year-perplexity-rank}
we visualize the release dates of the programming languages against
their median perplexity rank across models. Qualitatively, languages
at the low-perplexity end (\eg Go, C\#, Java) were released more
recently than several high-perplexity languages (\eg Shell, C). At
language granularity ($n=14$), Pearson correlation yields $r=-0.638$
($p = 0.0142$) and Spearman~$\rho=-0.719$ ($p = 0.00378$) between year
and median perplexity rank, \ie younger languages are associated with
lower median perplexity rank.

The high perplexity exhibited by C may be related to its long history;
throughout this, the programming landscape has changed significantly.
This also chimes with our earlier finding that C, even though
classified as strongly typed, ranked worse than scripting
languages.

\finding{%
  Median perplexity rank is negatively associated with the languages'
  release dates; younger languages are associated with lower median
  perplexity rank.}

\subsection{RQ3: How does code perplexity vary across LLMs on a multilingual code corpus?}
\label{sec:rq3}

We examined the correlation between our results and the results of the
two empirical studies of the related work
(Section~\ref{sec:related},~\citealp{izadi2024language,
  xu2022polycoder}). Because our study's setup differs from those used
in the other studies (\eg different datasets and model sizes), we
could not directly compare absolute perplexity values across studies.
Instead, we focused on the \emph{relative} language rankings produced
by each study, \ie how models order programming languages by
perplexity rather than their raw scores. For this we ran statistical
tests between our cross-model language ranking and the language ranking
produced by each model assessed in the other studies, by computing the
Spearman~$\rho$ and the Kendall~$\tau$ rank correlation coefficients,
which measure agreement between ordinal rankings without relying on
the absolute metric values.

Results are presented in Table~\ref{tab:rank-test}, sorted in
descending order by Spearman's~$\rho$ within each study. Next to each
coefficient we also provide the associated $p$-value. Our ranking
agrees significantly with two of the six models evaluated by Xu
\etal~\citeyearpar{xu2022polycoder}: GPT-Neo~2.7b and GPT-J~6b. None
of the three models evaluated by Izadi
\etal~\citeyearpar{izadi2024language} reaches statistical
significance, the strongest being InCoder. No coefficient in the table
is negative, so no ranking points in the opposite direction. These
tests, however, rest on the 9~and 8~languages we share with the two
studies, and at those sizes $\rho$ must exceed roughly
$0.67$ and $0.71$ respectively to reach $\alpha = 0.05$. The
non-significant rows, therefore, do not argue against agreement, but
bound the evidence.

In addition, we examined the correlation between the models of the
other studies, and identified statistical dependence in
PolyCoder~\citep{xu2022polycoder} between the rankings produced by
GPT-Neo~2.7b and GPT-J~6b, GPT-Neo~2.7b and GPT-NeoX~20b, GPT-J~6b and
GPT-NeoX~20b, and PolyCoder~2.7b and GPT-NeoX~20b. The first three
come as no surprise considering that the models come from the same
family. Consequently, the ordering of languages by perplexity appears
broadly consistent across studies and models, even where absolute
values are not comparable.

\begin{table}[t]
  \renewcommand{\arraystretch}{1.3}
  \caption{Correlation Between our Perplexity-based Language Ranking, Averaged
    Across all 18~Models, and Rankings by Models in Other Studies.
    Shaded rows are significant at $\alpha = 0.05$.}
  \label{tab:rank-test}
  \centering
  \begin{tabular}{l l r r r r}
    \hline
    \vspace{-0.3em}
    & & \multicolumn{2}{c}{Spearman} & \multicolumn{2}{c}{Kendall} \\
    Study & Model & \multicolumn{1}{c}{$\rho$} %
                  & \multicolumn{1}{c}{$p$} %
                  & \multicolumn{1}{c}{$\tau$} %
                  & \multicolumn{1}{c}{$p$} \\
    \hline
    \rowcolor{gray!15}
    \multirow{6}{*}{\citet{xu2022polycoder}}
    & GPT-Neo~2.7b~\citep{black2021gptneo}
      & $0.833$ & $0.005$ & $0.667$ & $0.013$ \\
    \rowcolor{gray!15}
    & GPT-J~6b~\citep{wang2021gptj}
      & $0.767$ & $0.016$ & $0.611$ & $0.025$ \\
    & GPT-NeoX~20b~\citep{black2022gptneox}
      & $0.567$ & $0.112$ & $0.444$ & $0.119$ \\
    & PolyCoder~2.7b~\citep{xu2022polycoder}
      & $0.383$ & $0.308$ & $0.333$ & $0.260$ \\
    & CodeParrot~\citep{tunstall2022natural}
      & $0.300$ & $0.433$ & $0.222$ & $0.477$ \\
    & Codex~\citep{chen2021codex}
      & $0.183$ & $0.637$ & $0.167$ & $0.612$ \\
    \hline
    \multirow{3}{*}{\citet{izadi2024language}}
    & InCoder~\citep{fried2023incoder}
      & $0.429$ & $0.289$ & $0.286$ & $0.399$ \\
    & UniXcoder~\citep{guo2022unixcoder}
      & $0.190$ & $0.651$ & $0.000$ & $1.000$ \\
    & CodeGPT~\citep{lu2021codexglue}
      & $0.190$ & $0.651$ & $0.000$ & $1.000$ \\
    \hline
  \end{tabular}
\end{table}

\begin{figure}[t]
  \centering
  \includegraphics[width=0.9\textwidth, keepaspectratio]{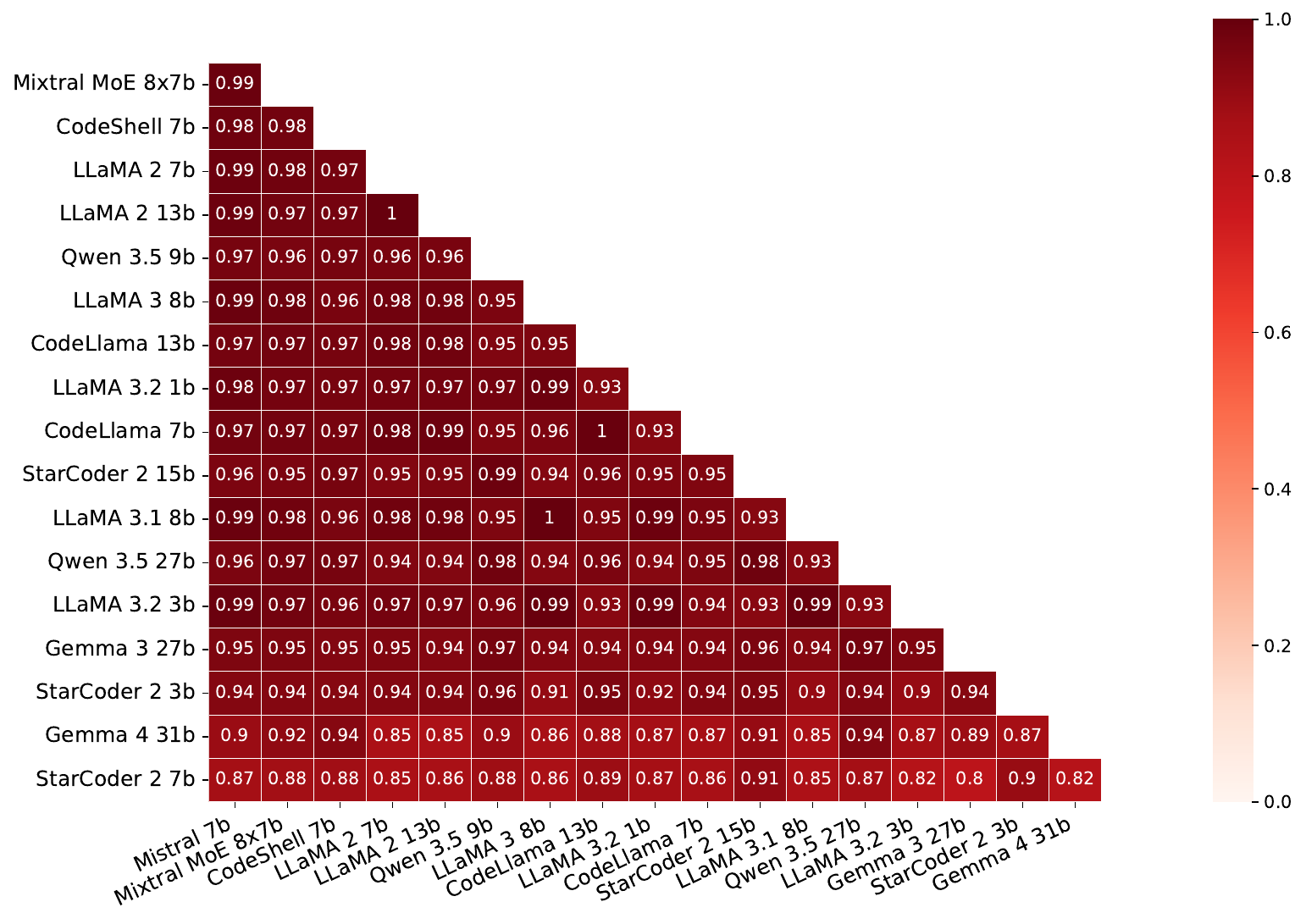}
  \caption{Pairwise correlation of LLMs based on their median perplexity.}
  \label{fig:perplexity-pairwise-correlation}
\end{figure}

Figure~\ref{fig:perplexity-pairwise-correlation} presents a pairwise
Pearson correlation coefficient heatmap of the LLMs computed from
their median perplexities across the programming languages. The LLMs
are sorted by mean correlation in descending order (axis labels run
from highest to lowest average agreement with all other models).
Correlations are high throughout: the median over the $153$ distinct
model pairs is $0.95$, and $125$ of them exceed $0.90$. Related models
appear as darker blocks, at $0.97$--$0.99$ (\eg Mistral and Mixtral
with the LLaMA line and CodeLlama, and StarCoder~2~15b with Qwen~3.5 or
CodeShell). The lightest region is the StarCoder~2~7b row, which has
the weakest average agreement with the rest ($0.86$) and contains the
three lowest cells in the figure: against Gemma~3~27b ($0.80$),
Gemma~4~31b ($0.82$), and LLaMA~3.2~3b ($0.82$). Gemma~4~31b comes next
($0.88$ on average), lightest against the LLaMA~2 pair and
LLaMA~3.1~8b ($0.85$ each). Only four of the $153$ pairs fall below
$0.85$, so even the weakest agreement stays strongly positive. In
short,
most models rank languages similarly, but not identically: changing
the model changes the numbers and can swap close pairs.

\finding{%
  Code perplexity depends on the chosen LLM: median perplexity
  profiles correlate strongly across model variants, yet absolute
  values and ordinal detail still shift between models, with related
  models agreeing more closely than unrelated ones. }

\begin{figure}[t]
  \centering
  \includegraphics[width=0.75\textwidth, keepaspectratio]{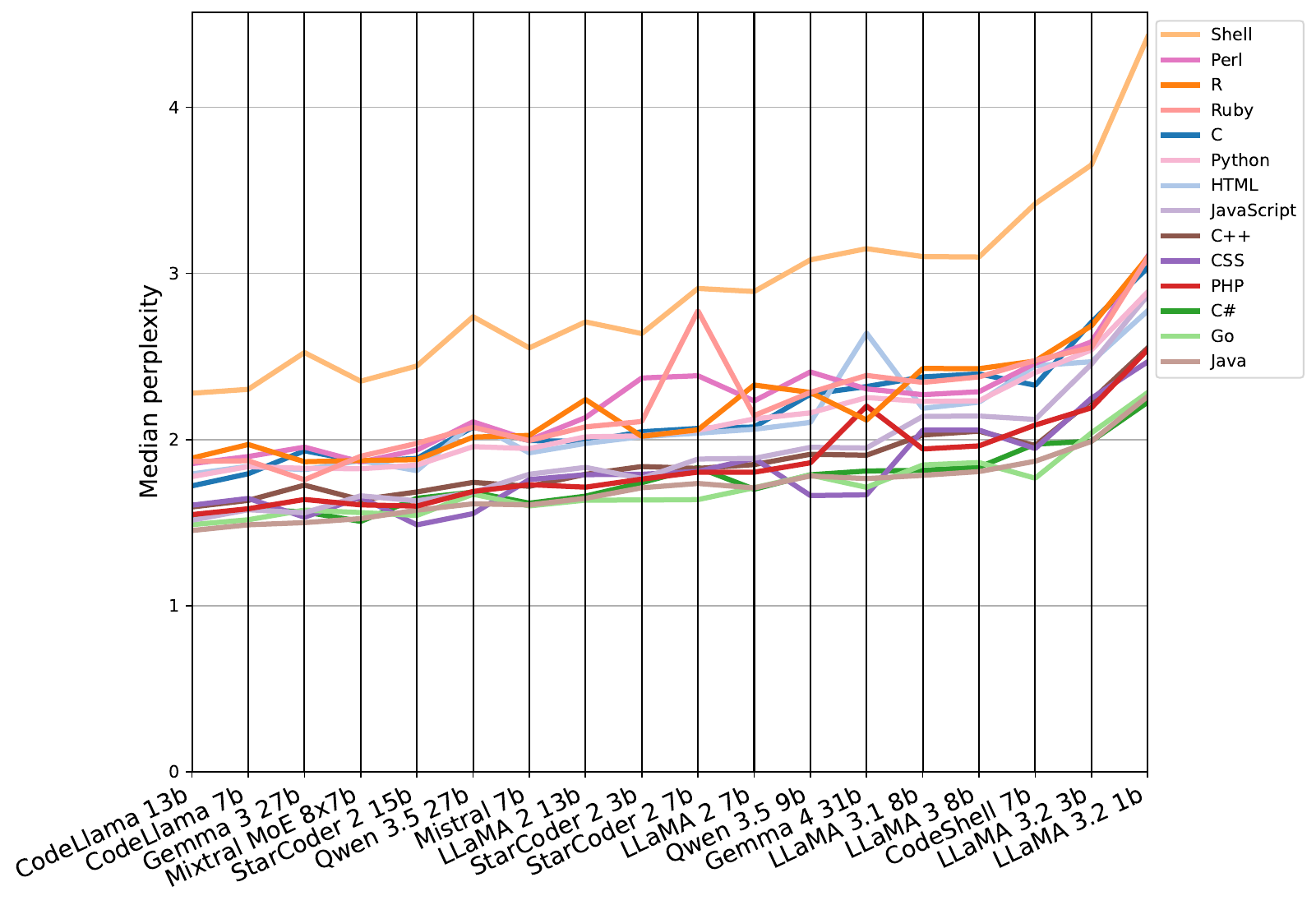}
  \caption{Perplexity across LLMs and programming languages.}
  \label{fig:perplexity-parallel-coordinates}
\end{figure}

\begin{figure}[t]
  \centering
  \includegraphics[width=0.75\textwidth, keepaspectratio]{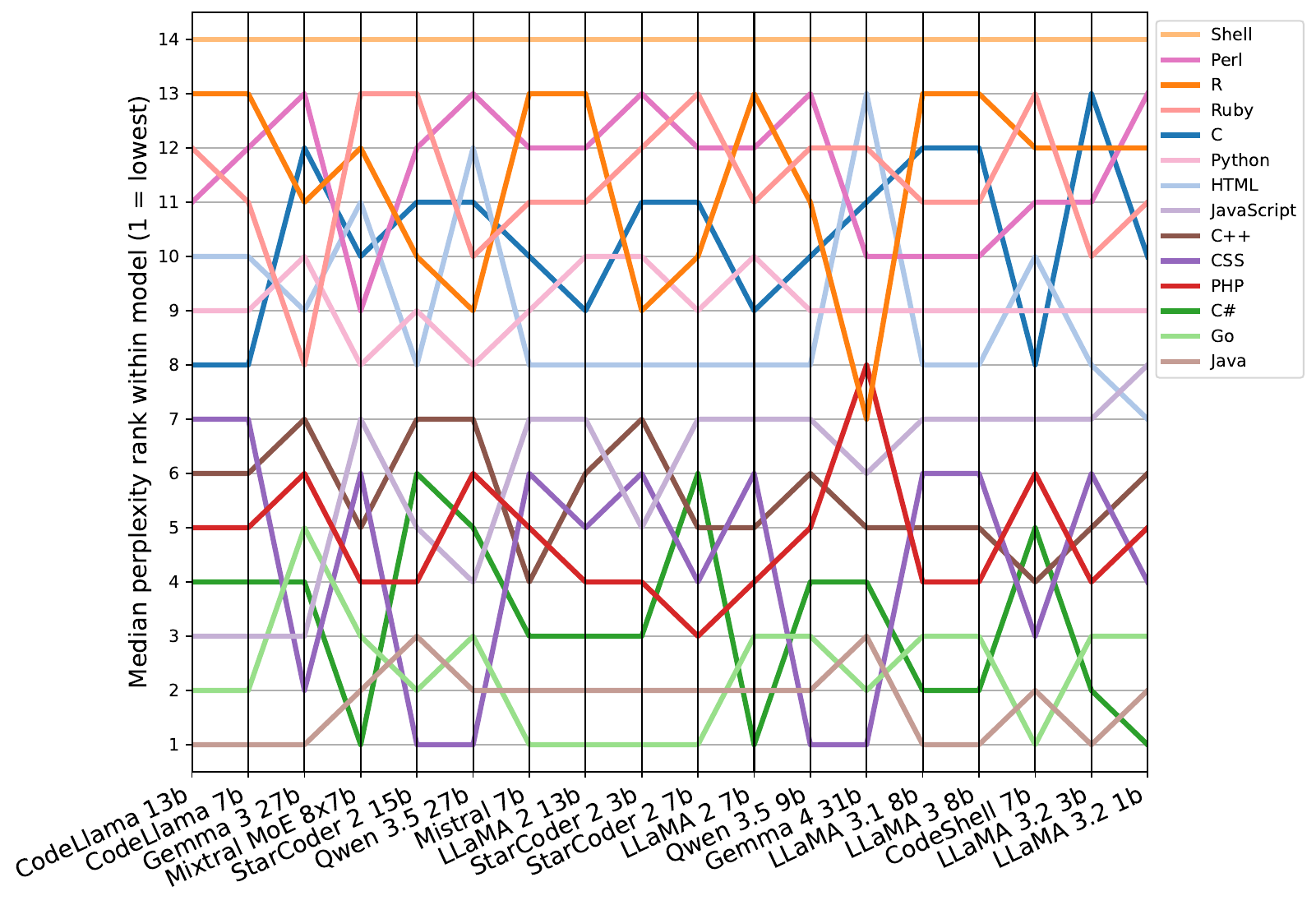}
  \caption{Language rank by median perplexity within each LLM
    ($1$ = lowest perplexity). Same models, order, and colors as
    Figure~\ref{fig:perplexity-parallel-coordinates}, with every model
    on a common scale.}
  \label{fig:perplexity-parallel-coordinates-rank}
\end{figure}

Furthermore, Figure~\ref{fig:perplexity-parallel-coordinates}
visualizes perplexity across LLMs and programming languages. Models
are sorted in ascending order of the median of their per-language
median perplexities. The figure demonstrates an agreement between
models in some clear-cut cases: Shell is universally high in
perplexity, whereas Java is low. All models share a single perplexity
axis there, so the height of a line mixes how hard a language is with
how high a model's perplexity runs overall; that mixture, not language
difficulty, is what makes every line rise from left to right.
Figure~\ref{fig:perplexity-parallel-coordinates-rank} repeats the plot
with the languages ranked by median perplexity within each model,
which puts all models on the same $1$--$14$ scale. Shell is then a
flat line at rank $14$: last in all $18$ models, and the only language
whose rank never varies. Java stays within ranks $1$--$3$, with Go
($1$--$5$) and C\# ($1$--$6$) beside it. Leaving Shell aside, the
languages separate into two bands. One band contains Java, Go, C\#,
PHP, CSS, C++ and JavaScript between ranks $1$ and $8$, while the
other band contains HTML, Python, C, Ruby, R and Perl between $7$ and
$13$. Nearly every crossing stays within a band: over all $18$ models
only two pairs ever swap across bands, PHP with R and JavaScript with
HTML, in one model each.

\finding{%
  Shell is universally high in perplexity,
  as opposed to Java which is universally low.
}

\subsection{RQ4: How do different evaluation datasets affect an LLM's measured code perplexity?}
\label{sec:rq4}

We tested whether language-level conclusions transfer when only the
\emph{evaluation corpus} changes. We held the model and tooling fixed,
and re-ran the same strided sliding-window computation on the
multilingual benchmark files released for the \emph{PolyCoder}
study~\citep{xu2022dataset}. We used the same context length, stride,
and toolchain parameters as for the analysis underlying
Figure~\ref{fig:perplexity-language-rank}
(Section~\ref{sec:ppl-implementation}), so differences in medians and
orderings are driven by the file set rather than the perplexity
configuration.

\begin{figure}[t]
  \centering
  \includegraphics[width=\textwidth,
  keepaspectratio]{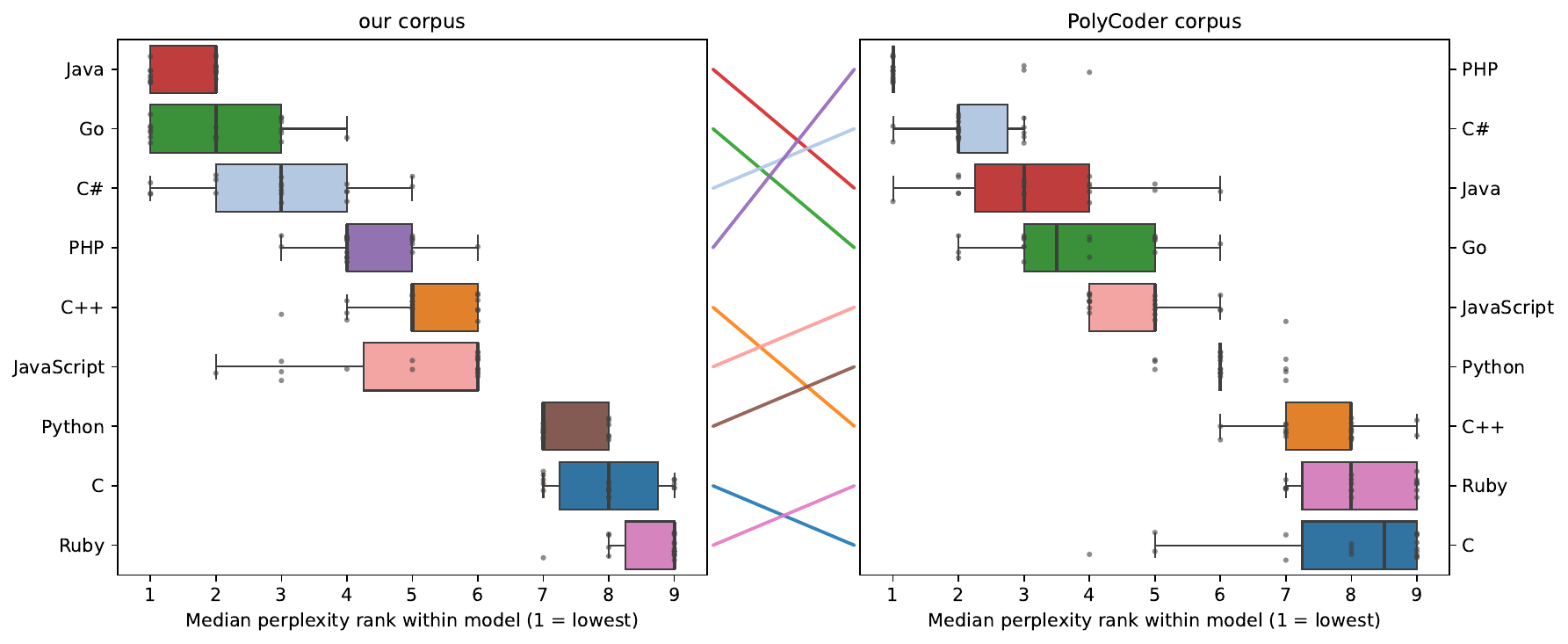}
  \caption{Language rank by median perplexity within each of the 18
    LLMs, on our corpus and on the \emph{PolyCoder} evaluation
    dataset, over the nine shared languages. Each point is one model.
    Each pane orders the languages by their median rank on its own
    corpus. The connecting lines join each language across the two.}%
  \label{fig:perplexity-polycoder-comparison}  
\end{figure}

Our sample of \emph{PolyCoder} files spans 12 languages; intersecting
them with the 14 languages in
Figure~\ref{fig:perplexity-language-rank} yields nine shared
languages: C, C\#, C++, Go, Java, JavaScript, PHP, Python, and Ruby.
Absolute perplexities are not comparable across two different file
sets, so we compare orderings. Within every model we rank the nine
shared languages by median perplexity on each corpus, which places both
on the same $1$--$9$ scale and removes the level difference between the
file sets. Figure~\ref{fig:perplexity-polycoder-comparison} shows the
resulting rank distributions over all $18$ models. Each pane orders the
languages by their median rank on its own corpus, and lines join each
language across the two, so the slope of a line is the change in
position.

Every model agrees with itself across the two corpora. We read the
$18$ per-model coefficients as dependent replications of one
comparison rather than as $18$ hypotheses: all of them are positive,
with a median of $0.73$, ranging from $0.30$ (CodeLlama~13b) to $0.85$
(LLaMA~3.2~1b and Mixtral~MoE~8x7b). Nine languages is a small sample,
so individual coefficients carry limited weight: $|\rho| > 0.67$ is
needed to reach $\alpha = 0.05$ at all, which $13$ of them do
uncorrected and $11$ under a Benjamini-Hochberg false-discovery
correction. Note that a family-wise correction (Holm-Bonferroni) is
not appropriate here, as the 18~models are near-duplicates of
one another, and the question is whether one shared phenomenon holds
rather than which model agrees.

The agreement is confirmed by the single comparison of the two
corpora's cross-model rankings, aggregating each language's rank over
the models by its median. That gives Spearman's $\rho = 0.75$
($p \approx 0.02$), with Kendall's $\tau = 0.51$ falling just short of the
threshold ($p \approx 0.06$) at the nine shared languages.

The residual disagreement is a reordering within groups rather than
across them. Four languages occupy the top on both corpora (Java, Go,
C\#, and PHP), three hold the middle ground (C++, JavaScript, and
Python), and two are at the bottom (C and Ruby), though no language
holds its exact position. The largest single move is PHP, from fourth
on our corpus to first on the \emph{PolyCoder} files; C++ drops from
fifth to seventh past JavaScript and Python, while C and Ruby just
trade the last two places.

\finding{%
  Under a fixed perplexity pipeline, the ordering of languages by
  median perplexity carries over to the PolyCoder benchmark: all 18
  models correlate positively across the two corpora, and their
  cross-model rankings agree at $\rho = 0.75$, so relative conclusions
  are broadly portable despite different absolute scores. }

\section{Discussion and Implications}
\label{sec:discussion}

Our study on code perplexity across programming languages, models, and
datasets has several implications for software engineering practice
and research. The confidence of LLM-based code assistance seems to
vary significantly by programming language. Strongly-typed languages
like Java, C\#, and Go consistently show lower perplexity, indicating
that developers working with these languages may experience more
accurate and reliable code suggestions. This aligns with previous
research by Husein \etal~\citeyearpar{husein2025llms}, who noted that
static type information provides clearer context for code prediction
models.

The perplexity variations across languages have direct implications
for development practices. Teams using languages with consistently
higher perplexity (Shell, C, R, Perl) might benefit from a more
cautious approach when adopting LLM-based tools, potentially
implementing stricter review processes for machine-generated code.
Conversely, projects using languages with lower perplexity could
benefit from streamlined workflows that leverage LLM suggestions more
extensively.

Perplexity's value as a confidence indicator provides practical
applications for code review and quality assurance. As demonstrated by
Sharma \etal~\citeyearpar{sharma2025assessing}, higher entropy
correlates with lower functional correctness in generated code.
Organizations could implement tiered review processes where
high-perplexity code blocks receive additional inspection, potentially
reducing error propagation in production systems. Such
confidence-based approaches could be particularly valuable for
critical codebases where reliability is paramount.

Our contribution is the open measurement pipeline and the empirical
basis for leveraging perplexity as a signal when using LLMs. Teams
could still use perplexity in an advisory way---\eg computing it on
files changed in a pull request and flagging them for extra review
when scores exceed a team-defined threshold, or showing file- or
region-level scores in a code review interface as a non-blocking hint
next to linters and tests. Doing so would require local choices of
model, budget, and cut-offs, and would benefit from calibration to the
deployment context, as for any confidence proxy
(Section~\ref{sec:limitations}).

Our finding that model choice shifts measured perplexity more strongly
than swapping the evaluation corpus (Section~\ref{sec:rq4}) has
significant implications for LLM development. It suggests that
\emph{architectural improvements might yield greater benefits than
  expanding training data}. The observed genealogical correlations
between model families (Mistral--Mixtral~MoE; Meta LLaMA line;
and CodeLlama--LLaMA~2;) indicate that
fundamental architectural decisions influence perplexity patterns
across model generations.

The observation that code comments modestly increase perplexity,
without shifting language rankings (Section~\ref{sec:rq2}), suggests
that natural-language text is somewhat less predictable, under the
same next-token model, than surrounding code tokens. Because the
effect is small and rank-preserving, it supports using perplexity as a
comment-inclusive, cross-language comparison metric without needing to
strip comments first, rather than as a signal about documentation
quality or prompting practice.

The relationship between language age and perplexity reveals an
interesting evolution in programming language design. More recent
languages with stronger typing systems and standardized syntax appear
more predictable by LLMs. Apparently, the same design principles that
enhance human readability and maintainability---clear structure,
consistent syntax, explicit typing---also improve machine
predictability. This insight could inform future language design
efforts, particularly for domain-specific languages.

The contrast between low perplexity in many strongly typed languages
and high perplexity in scripting languages in Section~\ref{sec:rq2},
leads to the hypothesis, consistent with Husein
\etal~\citeyearpar{husein2025llms}, that static types and fixed
interfaces contract the set of locally legal continuations, which
tends to lower next-token uncertainty for a model trained on large
code collections. The extent of possible continuations helps with the
``simple syntax'' puzzle in scripting: surface compactness is not the
same as low continuation entropy, because idiom diversity, dynamic
features, and glue-style automation (including markup and command
shells present in our sample) can leave many semantically different
next lines plausible, even when the grammar is small; training-data
mix and repository style can reinforce that pattern. To establish
causality, however, would require controlled corpora, syntax
manipulations, or ablations, which we do not perform here, and which
we mark as future work.

From a theoretical perspective, our results contribute to the broader
discussion on code naturalness~\citep{allamanis2018naturalness}. The
substantial rank agreement between our corpus and the \emph{PolyCoder}
files (Section~\ref{sec:rq4}) suggests that some predictability
differences between languages recur across corpora, while absolute
scores remain instance- and benchmark-dependent. This supports the
naturalness hypothesis while adding nuance: not all programming
languages are equally \emph{natural} from a modeling perspective.

Overall, our research findings can enhance several aspects of software
development. Organizations can make more informed decisions about
which LLM-based tools might be most effective for their technology
stack. Teams can implement confidence-based review processes that
allocate resources according to the expected reliability of
LLM-generated code. In addition, perplexity can serve as a
quantitative factor when evaluating potential language migrations,
particularly for teams seeking to maximize the effectiveness of
AI-assisted development. Finally, developers of code completion tools
can use perplexity to evaluate which models might perform best for
specific languages.

\section{Limitations}
\label{sec:limitations}

We present the risks resulting from our perplexity analysis
with respect to the internal and external validity of the study.

\subsection{Internal Validity}

The internal validity of the study is affected by particular steps
of our methods (Figure~\ref{fig:methods}).
The file filtering (Section~\ref{sec:file-preprocessing})
based on the language extensions by the \code{github-linguist/linguist} project
may have missed some relevant files whose extensions were not in the list.
Although we manually reviewed the extensions' list,
we may not have been able to correct all possible errors~\citep{PVK15}.

Another risk stems from the fact that
the files of the sample are of various sizes,
resulting in diverse numbers of prediction steps
during the perplexity computation of each file.
Although we did not find any association between file or vocabulary size
(in terms of tokens) and perplexity in our data (Section~\ref{sec:rq2}),
we cannot be certain whether an association generally exists.

The perplexity implementation we used is a close approximation of the
mathematical definition of perplexity. In theory, the complete history
is used in each prediction step. In practice, though, LLMs have a limited
context size, and the complete history may exceed that. As a
workaround, we used the strided method, and set a dynamic context size
so that all languages would be adequately represented
in the results (Section~\ref{sec:ppl-implementation}).

Another limitation to consider is that we did not calibrate the
perplexity metric before using it as confidence indicator. Recent work
by Spiess \etal~\citeyearpar{spiess2025calibration} discusses that
confidence metrics can benefit from calibration for decision-making
contexts. However, this limitation is mitigated by several factors.
First, our study focuses on comparative analyses between languages and
models rather than absolute confidence thresholds, and these relative
rankings are likely to remain valid even after calibration. Secondly,
our use of multiple LLMs for cross-validation provides additional
robustness to our findings beyond what calibration alone would offer.
Nevertheless, future work could explore calibration techniques such as
Platt scaling~\citep{platt1999probabilistic} to further refine the
relationship between perplexity scores and code correctness.

We also do not relate file-level perplexity to downstream outcomes
(\eg build or test failure rates, or defect labels)
on LLM-generated code in a single controlled run.
That would need a task-specific generation, execution, and scoring
pipeline---outside the scope of our intrinsic, human-authored corpus design.
Section~\ref{sec:intro} reviews why intrinsic measures are discussed
as potential proxies in the code LLM literature, and
Section~\ref{sec:discussion} considers heuristic uses of perplexity
in code review and quality workflows.
The present work does not add a direct empirical link from our file-level scores
to build, test, or defect outcomes;
closing that loop would require a dedicated generation-and-execution study,
which we treat as future work.

Finally, the inferential statistics we report have limitations common
to observational corpus studies. Rank-agreement tests (Spearman and
Kendall coefficients for comparing language orderings) operate at
small effective sample sizes---%
for example, only nine languages intersect our corpus and the
\emph{PolyCoder} benchmark---%
so $p$-values are coarse and sensitivity analyses would be
informative. Programming languages are not statistically independent
draws (they share semantics and syntax, ecosystems, tooling, and
corpora), so treating each language as one observation understates
dependence and can bias uncertainty downward. We also compare many
model pairs and study pairs without a formal multiple-testing
correction; the heatmap in
Figure~\ref{fig:perplexity-pairwise-correlation} is an exploratory map
and not a family-wide error-controlled test of hypotheses. Where a
family of comparable tests does arise, we correct within it:
Holm-Bonferroni across models for the comment and typing comparisons
of Section~\ref{sec:rq2}, and a false-discovery correction across
models in Section~\ref{sec:rq4}.

\subsection{External Validity}

A limitation of the main analyses in Sections~\ref{sec:rq1}
and~\ref{sec:rq2} is that they are computed on a single file sample
and corpus snapshot (Sections~\ref{sec:project-selection},~\ref{sec:file-preprocessing}),
even though they already aggregate results across all 18~models
(Section~\ref{sec:llm-selection}). The perplexity figures reported in
RQ1 and RQ2 are therefore tied to that file sample, and we would not
expect the same numbers to follow for an unrelated repository snapshot
without re-computation.

We assess this risk along two axes in RQ3 and RQ4
(Sections~\ref{sec:rq3},~\ref{sec:rq4}). RQ3 checks whether
language-level ordering is sensitive to \emph{model} choice, within
our own corpus: the pairwise correlation of median perplexity across
all 18~models is strongly aligned
(Figures~\ref{fig:perplexity-pairwise-correlation},~\ref{fig:perplexity-parallel-coordinates};
see the Finding in Section~\ref{sec:rq3}). The broad \emph{relative}
contrasts in language-level ordering (\ie which languages exhibit
lower versus higher median perplexity) therefore recur across
architectures, but absolute perplexities and some adjacent ranks in
the ordering still move with the model (cf.~Section~\ref{sec:rq3}).
RQ4 checks the complementary axis, whether ordering is sensitive to
\emph{corpus} choice, by re-running the same 18~models on the
\emph{PolyCoder} benchmark files: the rank-agreement tests on the nine
shared languages show largely consistent agreement of median orderings
(Section~\ref{sec:rq4} Finding), with some reordering among mid- and
lower-ranked languages, so \emph{relative} statements about language
ordering agree only in part when we compare the two file sets, and the
per-benchmark medians need not coincide. Taken together,
\emph{the RQ3 and RQ4 results suggest that language-level ordering may
  partially generalize when either the model or the evaluation dataset
  changes, whereas reported numeric perplexities remain tied to
  a particular model, corpus, and tooling configuration, and need to
  be recomputed before making outside comparisons.}

Generalizability concerns arise from the sample selection process employed
in project (Section~\ref{sec:project-preprocessing}) and
file selection (Sections~\ref{sec:file-preprocessing}).
We strived for a file sample
distributed evenly across programming languages and projects,
that at the same time would represent as many projects as possible.
Yet, the stability of the perplexity results could be strengthened
through cross-validation with multiple different samples.

Also, we selected the initial population of projects to be all
distributed with GPL licenses, to avoid evaluation overfitting (see
Section~\ref{sec:project-selection}), hence perplexity results may not
generalize to non-GPL-licensed projects. Furthermore, this exclusion
is verified only for LLaMA and, independently, for StarCoder~2, and even
then only partially: LLaMA's license filter applied to its dedicated GitHub
component, while other sources in its training mix (\eg \emph{CommonCrawl},
\emph{C4}) were not license-filtered and could still contain GPL code
incidentally. Moreover, we evaluate LLaMA~2,~3,~3.1,~3.2
(Section~\ref{sec:llm-selection}), not LLaMA itself, and their papers
make no comparable license-filtering claim. For the 17~models in our
study other than StarCoder~2, including these LLaMA variants, we
cannot confirm whether our GPL-licensed sample was actually excluded
from their training data. In addition, the correlation
output between perplexity and total/vocabulary size in
Section~\ref{sec:rq2} may be affected by our restricted sample size,
and should be repeated on larger samples to draw stable conclusions.

Our perplexity observations with respect to the language
characteristics in Section~\ref{sec:rq1} could be affected by the
representation of the languages in the training data. For instance,
mainstream languages could have more training data available impacting
model performance. We could only check this for LLaMA, the one model
whose training-project list we could retrieve
(Section~\ref{sec:project-selection}); this analysis of the effect of
the language distribution in the LLaMA training data on code
perplexity did not reveal any relationship between the two variables,
but we cannot be certain that such a relationship does not generally
exist for LLaMA or for the other 17~models in our study.

\section{Conclusion}
\label{sec:conclusion}

We investigated code perplexity through various LLMs on a multilingual
code corpus. Perplexity~\citep{jelinek:1977, huyen2019evaluation} is a
simple, efficient, versatile, and straightforward intrinsic metric
that can be applied universally across LLMs and
tasks~\citep{huyen2019evaluation, xu2022polycoder}. Lower perplexity
has been found to decrease error rate~\citep{bahl1983likelihood} and
improve end-task performance~\citep{liu2019roberta, xu2022polycoder,
  nijkamp2023codegen}, offering an effective and efficient proxy for
downstream metrics. Still, limited research has investigated LLM
performance from this perspective.

To fill this gap we conducted an empirical analysis of a sample of
2254 files coming from 881 GPL-licensed projects from GitHub, spanning
14 programming languages. Through our analysis we evaluated code
perplexity variation across languages, models, and datasets, and
compared our results with two other empirical
works~\citep{izadi2024language, xu2022polycoder}. We found that
strongly-typed languages show lower code perplexity than dynamically
typed languages. Scripting languages demonstrate higher perplexity,
with Shell appearing universally high as opposed to Java which appears
universally low in perplexity. Although code comments modestly increase
perplexity, the language ranking based on perplexity is barely
affected. We did not identify any correlation between vocabulary or
file size (in tokens) and code perplexity. Lastly, median code
perplexity depends on the employed LLM, whereas language rankings
based on median perplexity largely carry over when re-evaluating the
same model on the \emph{PolyCoder} benchmark files, aside from some
reordering among mid- and lower-ranked languages.

Our findings allow LLM researchers, developers, and users to know how
beneficial the use of an LLM assistant can be in a software project
depending on its language as well as what LLM to choose. Also, they
can understand how a model's confidence is affected by different
project, language, and code characteristics such as language verbosity
and type system, vocabulary size, and code comments.

Future research could investigate further the impact of vocabulary and
file size on perplexity. Some supplementary questions worth
researching concern how project, developer, and additional language
characteristics may affect code perplexity, and how code perplexity
correlates with LLMs' downstream performance metrics.







\bibliography{codepred}

\end{document}